\documentclass[a4paper,useAMS,usenatbib]{mnras}
\usepackage{amsmath}
\usepackage{amssymb}
\usepackage{float}
\usepackage{graphicx}
\usepackage{hyperref}

\newcommand{\vect}[1]{\bmath{#1}}
\newcommand{\upright}[1]{\mathrm{#1}}
\newcommand{\unit}[1]{\, \mathrm{#1}}
\newcommand{\subscript}[1]{_{\mathrm{#1}}}

\newcommand{\changed}[1]{{#1}}

\title[A small-scale dynamo in feedback-dominated galaxies]{A small-scale dynamo in feedback-dominated galaxies as the origin of cosmic magnetic fields - I. The kinematic phase}
\author[M. Rieder, R. Teyssier]{Michael Rieder\thanks{Contact e-mail: \href{rieder@physik.uzh.ch}{rieder@physik.uzh.ch}} and Romain Teyssier\\
Institute for Computational Science\\
Centre for Theoretical Astrophysics and Cosmology\\ 
Universit\"at Z\"urich, 8057 Z\"urich, Switzerland}

\begin{document}
\maketitle

\begin{abstract}
\changed{The origin and evolution of magnetic fields in the Universe is still an open question. Their observations in galaxies suggest strong magnetic fields already at high redshift as well as at present time. However, neither primordial magnetic fields nor battery processes can account for such high field strengths, which implies the presence of a dynamo process with rapid growth rates in high-redshift galaxies and subsequent maintenance against decay.} \par \changed{We investigate the particular role played by feedback mechanisms in creating strong fluid turbulence, allowing for a magnetic dynamo to emerge. Performing magnetohydrodynamic simulations of isolated cooling gas halos, we compare the magnetic field evolution for various initial field topologies and various stellar feedback mechanisms. We find that feedback can indeed drive strong gas turbulence and dynamo action. We see typical properties of Kolmogorov turbulence with a $k^{-5/3}$ kinetic energy spectrum, as well as a small-scale dynamo, with a $k^{3/2}$ magnetic energy spectrum predicted by Kazantsev dynamo theory. We also investigate simulations with a final quiescent phase. As turbulence decreases, the galactic fountain settles into a thin, rotationally supported disk. The magnetic field develops a large-scale, well-ordered structure with even symmetry, which is in good agreement with magnetic field observations of nearby spirals. Our findings suggest that weak initial seed fields were first amplified by a small-scale dynamo during a violent, feedback-dominated early phase in the galaxy formation history, followed by a more quiescent evolution, where the fields have slowly decayed or were maintained via large-scale dynamo action.}
\end{abstract}


\section{Introduction}

Large-scale magnetic fields are measured with strengths of up to several $\mu$G in nearby galaxies \citep{1996ARA&A..34..155B},
and possibly even higher field strength have been detected in earlier galaxies at high redshift \citep{2008Natur.454..302B}. 
The preferred theory to explain their origin is based on
the early generation of seed  fields at the epoch of cosmic re-ionisation, through the microscopic process known as 
"Biermann battery" \citep{1950ZNatA...5...65B,2013PhRvL.111e1303N},
later amplified during the galaxy formation era, through the large-scale galactic dynamo process \citep{1970ApJ...160..383P, 2005PhR...417....1B}.
The Biermann battery is likely to generate fields at the level of $10^{-20}$~G, leaving to the galactic dynamo process more than 14 orders of magnitude
of field amplification during the 10~Gyr of cosmic evolution. The challenge for galactic dynamos is even more severe, if one considers that strong fields are already in place
at high redshift \citep{2002RvMP...74..775W}, and are probably even stronger than they are today \citep{2008Natur.454..302B}.

Successful theoretical models for large-scale galactic dynamos report exponential growth rates of the order of $\Gamma \simeq 0.01$ to $0.1 \Omega$,
where $\Omega$ is the galactic disk rotation rate \citep{pariev2007magnetic}. 
For typical, present day spirals, this translates into e-folding amplification time scale of roughly 1~Gyr, 
making the task of amplifying the field over 14 orders of magnitude virtually impossible. One noticeable exception is the cosmic-ray-driven dynamo 
proposed by \cite{1992ApJ...401..137P} and simulated by \cite{2004ApJ...605L..33H}, leading to a measured growth rate $\Gamma \simeq \Omega$, 
although the numerical experiment was performed over only a relatively limited time, since the reported magnetic field amplification was over only 
3 orders of magnitude \citep{2004ApJ...605L..33H}. On the theoretical side, classical mean field dynamos are plagued by the catastrophic
$\alpha$-quenching effect, leading to very low saturated values for the large-scale magnetic field \citep{1992ApJ...396..606K,1992ApJ...393..165V}, 
owing to the strict conservation of magnetic helicity in a closed system. A possible solution to this problem is the effect of galactic winds, that could 
drag the magnetic field lines outside of the dynamo-active disk, therefore alleviating the aforementioned quenching issue \citep{2013MNRAS.429.1686D}.

The theory of galaxy formation has significantly evolved over the past decade, with the ever increasing role of feedback processes \citep{2012MNRAS.tmp.2970S} and their associated galactic winds
\citep{2006MNRAS.373.1265O}, together with the dominant accretion mechanism through cold streams \citep{2005MNRAS.363....2K,2008MNRAS.390.1326O,2009Natur.457..451D}. 
On the observational side, galactic winds are indeed ubiquitous in star bursting local galaxies \citep{1999ApJ...513..156M}, but also in many 
``normal'' high redshift galaxies \citep{2010ApJ...717..289S}. One of the most spectacular observational constraints on galaxy formation theories was obtained by matching the
stellar mass of the central galaxies to their parent halo mass \citep{2013ApJ...770...57B,2013MNRAS.428.3121M}. This has led theorists to consider much stronger feedback processes,
in order to regulate star formation throughout cosmic time, especially at high redshift, when the star formation efficiency was so low
\citep{2013ApJ...770...25A,Hopkins:2014ua,2014MNRAS.444.2837R,Wang:2015dc}.

In this rather violent, feedback-dominated scenario, dwarf galaxies play a very important role. They are the dominant galaxy population at high redshift, probably responsible for the cosmic re-ionisation
\citep{2014ApJ...788..121K}. They are also the progenitors of the Milky Way satellites, which are useful laboratories to test our current galaxy formation paradigm. For the latter,
violent feedback mechanisms have also been invoked to explain the absence of cusp in the dark matter density profile, and the presence of a dark matter core in low surface 
brightness galaxies \citep{2001ApJ...552L..23D}. Cosmological simulations of dwarf galaxies have been performed with strong feedback recipes, confirming in this case the formation of a dark matter core 
\citep{2010Natur.463..203G,2012MNRAS.422.1231G}. Recently, we have also performed idealised simulations of an isolated, cooling gaseous dwarf halo, obtaining, in this well-controlled numerical 
experiment, the formation of a dark matter core \citep{2013MNRAS.429.3068T}. The dark matter core formation mechanism is now well understood \citep{2012MNRAS.421.3464P}. 
It is due to repeated, energetic feedback events
due to many supernovae explosions, leading to violent oscillations of the gravitational potential, due to the large gas mass variations within the central kilo parsec of the galaxy. A possible observational 
signature of this effects is a typical, oscillatory star formation history, mimicking a breathing mode in the gas distribution \citep{2014MNRAS.441.2717K}.

In this paper, we want to study the impact of a strong feedback scenario on the growth of magnetic fields in dwarf, as well as in larger galaxies. The velocity field on both small and large scales,
resulting from repeated giant feedback events, can have a direct influence on the growth of the magnetic energy. Indeed, 
supernovae explosions in the Milky Way have been considered for quite a long time 
as a source of helical gas motions, promoting a large-scale $\alpha$-dynamo in the Galaxy \citep{1992ApJ...391..188F}. The Milky Way is however a rather quiescent galaxy, with moderate supernovae activity. In this paper, we are considering feedback-dominated galaxies, with high star formation rates and violent turbulent motions, together with large-scale galactic fountains or winds. 

Several simulations including magnetic fields have been performed recently in the context of galaxy formation \citep{2009ApJ...696...96W,2010A&A...523A..72D}. 
These simulations, based on the popular ``cooling halo'' numerical set-up, 
have achieved only moderate magnetic field amplification. The important property of these simulations is the absence of feedback \citep{2009ApJ...696...96W}, or the relative weakness of the feedback recipe used at that time \citep{2010A&A...523A..72D}. 

A first  exception is the simulation reported in \cite{beck2012origin}, based on a MHD version of the SPH code GADGET with divergence cleaning. They observed a fast exponential growth of the magnetic field,  which they attribute to a small-scale dynamo. Feedback processes were included through an effective Equation-of-State (EoS), without any explicit source of turbulence in these relatively smooth, thermally-supported flows. These authors however reported very strong growth rates, with e-folding times as small as 10~Myr, although analytical estimates based on small-scale dynamo theory predicted e-folding times closer to 100~Myr. 

A second exception is the recent simulation reported in \cite{2013MNRAS.432..176P}, using the new Magneto-Hydrodynamics (MHD) solver developed for the AREPO code \citep{2011MNRAS.418.1392P}, where strong magnetic field amplification has also been observed, although, here again, stellar feedback effects were not considered explicitly, but only indirectly as a modified thermal EoS, leading to the formation of relatively smooth, two-dimensional flows, in which dynamo amplification is in principle notoriously difficult to obtain.

In the present paper, we will use a similar set-up as in all those previous studies, namely a cooling isolated gaseous halo, considering simulations with (but also without) strong stellar feedback. We will use the Adaptive Mesh Refinement code RAMSES \citep{2002AA...385..337T}, adopting the ``Constrained Transport'', strictly divergence-free-preserving, MHD solver presented in \cite{2006JCoPh.218...44T} and in \cite{2006AA...457..371F}. The paper is organised as follows: In \autoref{chap:num-methods}, we will present our numerical methods, both in terms of galaxy formation physics and magnetic field modelling. In \autoref{chap:IC}, we describe our initial conditions for the isolated, {\it magnetised} cooling halo. In \autoref{chap:results}, we present our main results, outlining the difference between the feedback and the no-feedback cases. Finally, in \autoref{chap:discussion}, we discuss the implications of our results in the context of galactic dynamo theory, as well as possible further studies to confirm and broaden our findings.

\section{Numerical methods}
\label{chap:num-methods}

We have performed MHD simulations of isolated, cooling haloes, using the Adaptive Mesh Refinement (AMR) code RAMSES \citep{2002AA...385..337T}. 
These simulations feature a collisionless fluid (for dark matter and stars) and a magnetised gaseous component, coupled through gravity.
In this section, we describe the simulation technique used to follow the evolution of our isolated halo. First, we describe in details our AMR implementation for solving the ideal MHD equations, together with simple test cases to show that it works as intended in the context of galactic dynamo. We then describe the adopted galaxy formation physics, such as gas cooling, metal enrichment, star formation and stellar feedback, leading to what we believe to be a realistic model of the interstellar medium (ISM).

\subsection{Ideal MHD solver}
The equations that we solve are the ideal MHD equations (written here without gravity and cooling source terms for the sake of simplicity)
\begin{align}
\partial_{t} \rho + \nabla \cdot (\rho \vect u) &= 0 \\
\partial_{t} (\rho \vect u) + \nabla \cdot ( \rho \vect u \vect u^{T} - \vect B \vect B^{T} + P \subscript{tot} ) &= 0 \\
\partial_{t} E + \nabla \cdot \left[ (E + P \subscript{tot}) \vect u - (\vect u \cdot \vect B) \vect B \right] &= 0 \\
\partial_{t} \vect B - \nabla \times \left( \vect u \times \vect B \right) &= 0
\label{induction-eqn}
\end{align}
where $\rho$ is the gas density, $\rho \vect u$ is the momentum, $\vect B$ is the magnetic field, $E = \frac 1 2 \rho \vect u^{2} + \varepsilon + \frac 1 {2} \vect B^{2}$ is the total energy, and $\varepsilon$ is the internal energy. The total pressure is given by $P_{tot} = P + 
\frac{1}{2} \vect B^{2}$ where we assume a perfect gas equation of state $P = (\gamma - 1) \varepsilon$. The system of equations is completed by the soloinoidal constraint
\begin{equation} \nabla \cdot \vect B =0 .\end{equation}

The code is grid-based with a tree-based adaptively refined mesh. The equations are solved using the second-order unsplit Godunov scheme based on the MUSCL-Hancock method. We chose the HLLD Riemann solver with the MinMod slope limiter for the hydro variables which are cell-centred. The magnetic field on the other hand is treated as a face-centered variable. This allows the use of the Constrained Transport (CT) method to advance the induction equation (\autoref{induction-eqn}) in time, which preserves the divergence of the magnetic field to the numerical precision level \citep{2006JCoPh.218...44T}. The CT method involves a spatial interpolation of the EMF on the cell edges for the predictor step and solving a 2D Riemann problem for the corrector step. For the 2D problem, we use the HLLD solver as well and for the magnetic field in general, the MonCen slope limiter.

Boundary conditions were chosen to allow for free outflow. For the 5 hydro variables, this is done by imposing a vanishing gradient at the domain boundary (zero-gradient method). The same can be applied to the transverse magnetic field component parallel to the boundary face $B_{\|}$, but would cause a non-zero divergence of the magnetic field if applied to the normal component $B_{\bot}$ which is perpendicular to the face. Instead, we use a linear interpolation for $B_{\bot}$ so that $\nabla \cdot \vect B = 0$. Note that this method can cause an inward Poynting flux which transports magnetic energy from the outside into the computational domain. Since the magnetic field at the border is many orders of magnitude weaker than the average, this does not contribute significantly to the overall magnetic energy evolution \citep[see][]{2010A&A...523A..72D}.

Special care needs to be taken also when refining and de-refining cells, in order to enforce the $\nabla \cdot \vect B = 0$ constraint, when interpolating the magnetic field. A solution to this problem within the CT framework has been proposed by \cite{2001JCoPh.174..614B} and \cite{2002JCoPh.180..736T}, and we adopt it here for newly refined cells, but also for temporary ghost cells used to set  proper boundary conditions at coarse-fine level boundaries.

In the context of galactic dynamos, it is worth mentioning that our code has been tested extensively against well-known flows triggering fast dynamos, such as in the ABC flow \citep{1986GApFD..36...53G,1995stf..book.....C} or in the Ponomarenko dynamo \citep{1973JAMTP..14..775P}. We have shown in \cite{2006JCoPh.218...44T} that our numerical scheme for the ideal MHD equations 
was in fact slightly resistive, with, for a regular Cartesian grid, a numerical magnetic Reynolds number roughly inversely proportional to the square of the number of grid points per box length. 
This scaling is to be expected for second-order schemes and smooth flows. In the context of AMR and highly complex, turbulent flows, determining the exact effective numerical Reynolds number of the simulated flow is  impossible. Qualitatively, though, it is important to bear in mind that magnetic reconnection and other diffusive processes occur in the simulation at a typical scale probably very close to the grid scale. This scale plays a very important role in dissipating the kinetic energy of the turbulence, and also controls the magnetic energy losses due to reconnection or (numerical) Ohmic dissipation. 

\subsection{Cooling and star formation}

In addition to solving the ideal, self-gravitating MHD equations, we also include many physical processes relevant to galaxy formation. One of the key physical ingredient is gas cooling, which leads the hot,
initially hydrostatic halo gas to loose pressure support and to condense in the centre as a centrifugally supported disc. When this atomic gas of $10^4$~K is allowed to cool even more due to fine-structure metal line cooling or molecular cooling, the disc fragments into dense clumps, leading to the formation of a turbulent, multiphase medium. 
To model gas cooling, we use standard H and He cooling processes with an additional contribution from metals based on \cite{1993ApJS...88..253S} for temperatures above $10^{4}$ K and metal fine-structure cooling below $10^4$~K, as in \cite{1995ApJ...440..634R}. The metallicity, denoted as $Z$, is modelled as a passive scalar, representing the mass fraction of atoms heavier than Helium in the gas. It is initialised to $Z_{\rm ini} = 0.05 Z_{\odot}$ in the halo, mimicking molecular Hydrogen cooling in a zero metallicity gas. The metallicity is increased further by supernova feedback events, using a metal yield of $y = 0.1$. 

Our refinement strategy is based on a quasi-Lagrangian approach: each cell is refined if it contains more than 8 dark matter particles or if its baryonic mass (including gas and star particle mass) exceeds $8 \times m_{\rm res}$, where $m_{\rm res}$ is the adopted mass resolution of the simulation. Refinement are performed recursively, on a cell--by--cell basis, until the adopted maximum level of refinement is reached (noted $\ell_{\rm max}$).
It is crucial for astrophysical simulations to resolve spatially the Jeans length \citep{1997ApJ...489L.179T}. Requiring that the Jeans mass is resolved by at least $64$ mass resolution elements, $M_{\rm J}= 64 m_{\rm res}$, and adopting a realistic minimum temperature for the gas, noted $T_{\rm J}$, one can compute the corresponding Jeans length, and require it to be resolved by 4 cells, $\lambda_{\rm J}= 4 \Delta x_{\rm min}$. We can then determine the maximum required level of refinement corresponding to the adopted mass resolution $m_{\rm res}$. To prevent the gas from accumulating and locally violating the Jeans length criterion, we also add an artificial pressure floor,
\begin{equation}
P_{J} = (4\Delta x\subscript{\rm min})^{2} \frac G {\pi \gamma} \rho^{2}
\end{equation}
so that the gas density will never significantly exceed a typical value $n_{\rm J}$ given by the relation $k_{\rm B}T_{\rm J}=P_{\rm J}(n_{\rm J})/n_{\rm J}$.

Stars are treated as collisionless particles which are created stochastically from the gas according to a Schmidt law \citep[as in][]{2006A&A...445....1R}
\begin{equation}
\dot{\rho}_{*} = \epsilon_{*} \frac{\rho\subscript{gas}}{t\subscript{ff}}
\end{equation}
if the local density $\rho_{\rm gas}$ is above a threshold density $\rho_{*} = n_{*} \unit{m_{H}}$. In this paper, we always choose the star formation threshold density to be equal to the previously defined 
Jeans density $n_{\rm J}$.
The star formation efficiency per free-fall time is always set to $\epsilon_{*} = 0.01$; this value is based on observations of nearby molecular clouds \citep{2007ApJ...654..304K}. Creation of stellar particles is a local random Poisson process with Poisson parameter $\lambda = \rho_{*}\Delta x^{3}\Delta t / m_{*}$ where $\Delta t$ is the simulation time-step and
\begin{equation}m_{*} = n_{*} \left(\Delta x\subscript{\rm min}\right)^3\end{equation}
the mass of the resulting stellar particle, which is equal to the smallest cell mass at the density threshold. For each simulation, precise numbers for all these parameters are given in \autoref{haloic}.

\subsection{Stellar feedback}
\label{sect:feedback}

In this paper, we explore the consequences of strong feedback scenarios on the amplification of the magnetic field in dwarf and Milky-Way-sized galaxies.
The proper modelling of stellar feedback mechanisms, such as supernovae explosions, photo-ionised bubbles or infrared radiation in dusty environment is subject to intensive research 
throughout the ISM and galaxy formation literature. Understanding in details these various processes goes far beyond the scope of this paper. Our goal is merely to use various phenomenological
recipes to model such feedback mechanisms very crudely, and produce dynamical properties that we believe are relevant for high-redshift galaxies, the most important one being the gas velocity field,
highly turbulent, explosive and fountain-like, which could result in a fast magnetic dynamo. 

\subsubsection{Supernovae feedback}

For this purpose, we used a numerical model for supernovae feedback developed in the context of dwarf galaxies evolution, and that turned out to 
lead to the formation of a dark matter core \citep{2013MNRAS.429.3068T}. 
The main ingredient is the use of a non-thermal energy variable, and its associated pressure, treated as a 
passively advected scalar quantity $e\subscript{turb} = \rho \epsilon\subscript{turb}$, which represents various small-scale, 
non-thermal energies released by supernovae (e.g. turbulence, magnetic fields or cosmic rays). The evolution of this non-thermal energy is specified by
\begin{equation}
\rho \frac{D \epsilon\subscript{turb}}{D t} = \dot{E}\subscript{inj} - \frac{\rho \epsilon\subscript{turb}}{t\subscript{diss}}.
\end{equation}
where the dissipation time scale is fixed to $t_{\rm diss}=20$~Myr and the energy injection per supernovae is set by
\begin{equation}
\dot{E}\subscript{inj} = \dot{\rho}_{*} \eta \subscript{SN} \cdot 10^{50} \unit{erg/M}_{\odot}
\end{equation}
where the mass fraction
in massive stars is set to $\eta_{\rm SN}=0.1$ and the local star formation rate $\dot{\rho}_{*}$ is set by our adopted Schmidt law. For details about the implementation, we refer to \cite{2013MNRAS.429.3068T}

\subsubsection{Radiation feedback}

Because supernovae are probably not energetic enough to trigger strong winds in Milky-Way-sized galaxies, it has been proposed recently 
to consider stellar radiation as an additional feedback mechanism \citep{2010ApJ...709..191M}.
Interstellar dust indeed absorbs UV photons, much of it being subsequently re-emitted as thermal radiation in the infrared band. 
This radiation will transfer momentum to the gas through radiation pressure
\citep{2010ApJ...709..191M,2012MNRAS.tmp.2654H,2013ApJ...770...25A,2014MNRAS.444.2837R}
In this paper, we consider radiation feedback only for Milky-Way-sized galaxy simulations. We use here again a very crude model to capture the energy from the stellar UV radiation,
using a simple escape probability model as
\begin{equation}
E \subscript{UV} = E \subscript{rad} \left[ 1- \mathrm{exp}(-\kappa\subscript{UV}\rho\subscript{dust}\Delta x) \right]
\end{equation}
with the dust mass density is assumed to be $\rho\subscript{dust} = Z \rho\subscript{gas}$ (here, Z denotes the gas metallicity). The dust opacity at $0.1 \unit{\mu m}$ is taken to be $\kappa\subscript{UV} = 1000 \unit{cm^{2}/g}$ \citep{2007ApJ...657..810D}, and the total energy released during the first 10 Myr of a $10 \unit{M_{\odot}}$ progenitor $E \subscript{rad} = 10^{52} \unit{erg/M_{\odot}}$. The same cell is then assumed to absorb the energy in the infrared
\begin{equation}
E \subscript{IR} = E \subscript{UV} \left[ 1- \mathrm{exp}(-\kappa\subscript{IR}\rho\subscript{dust}\Delta x) \right]
\end{equation}
where $\kappa\subscript{IR}$ is the dust opacity in the IR band, which is a free parameter in our feedback implementation, usually around 10~cm$^2/$g \citep{2007ApJ...657..810D,2003A&A...410..611S}. The energy $E \subscript{IR}$ is added in the non-thermal energy 
equation for $\dot{E}\subscript{inj}$ in the supernova feedback so that it contributes to $\epsilon\subscript{turb}$. Details about the implementation can be found in \cite{2014MNRAS.444.2837R}.
Here again, we would like to stress that our goal is not to study in great details realistic feedback mechanisms, but rather to generate galactic velocity fields in qualitative agreement with 
high-redshift galaxies and their associated strong outflows, in the context of galactic dynamos.

\section{Initial conditions}
\label{chap:IC}
We have performed a series of non-cosmological simulations of isolated halos in hydrostatic equilibrium, varying the initial set-up with two different halo sizes (a typical dwarf and a typical Milky-Way) and two different initial magnetic field topologies (dipole and quadrupole), in addition to the various options for stellar feedback that we have discussed in the previous section. We will now describe more precisely our initial set-up, and a summary of the various run parameters is given in \autoref{feedbackic}. \changed{We refer the interested reader to \cite{2013MNRAS.429.3068T} for a more detailed analysis of the non-magnetohydrodynamic properties in the dwarf halo case.}

\subsection{Initial halo}
Our initial halo follows a \cite{1997ApJ...490..493N} (hereafter NFW) density profile with a concentration parameter $c=10$. The smaller of the two halos, representative of a typical dwarf galaxy (we use the acronym DW), has a circular velocity of $V_{200} = 35 \unit{km/s}$ and a virial mass of $M_{200} = 1.4 \times 10^{10} \unit{M_{\odot}}$, both measured at the virial radius $R_{200} = 50 \unit{kpc}$ . The halo is truncated at $112.5 \unit{kpc}$, resulting in the total enclosed mass of $2 \times 10^{10}\unit{M_{\odot}}$. This is essentially the same set of parameters we used in the hydrodynamic simulations of \cite{2010A&A...523A..72D} and \cite{2013MNRAS.429.3068T}.

The larger halo is chosen to be a typical Milky Way galaxy (we use the acronym MW), where we increased the circular velocity to $V_{200} = 160 \unit{km/s}$, corresponding to a virial radius of $R_{200} = 230 \unit{kpc}$ and a virial mass of $M_{200} = 1.3 \times 10^{12} \unit{M_{\odot}}$. It is truncated at $514 \unit{kpc}$, resulting in the total enclosed mass of $2 \times 10^{12}\unit{M_{\odot}}$. 

In all other aspects, both initial configurations follow the same prescription as in \cite{2013MNRAS.429.3068T}. We consider a gas fraction equal to the universal mean value $f \subscript{gas} = 15 \%$, and the gas density is also following a NFW profile. The gas temperature is initialised by solving the hydrostatic equilibrium equation. The gaseous halo is set in slow rotation around the z-axis, using the angular momentum profile from cosmological simulations and a spin parameter $\lambda = 0.04$. The dark matter halo is sampled by $10^{6}$ dark matter particles, whose initial positions and velocities were computed with the density-potential pair approach of \cite{2004ApJ...601...37K} and \cite{2006MNRAS.367..387R}. The stability of the resulting gas-dark matter equilibrium was shown in \cite{2013MNRAS.429.3068T} to be sufficiently good for our present purpose.

\subsection{Initial magnetic field}

\begin{figure*}
\centering
\includegraphics[width=0.667\linewidth]{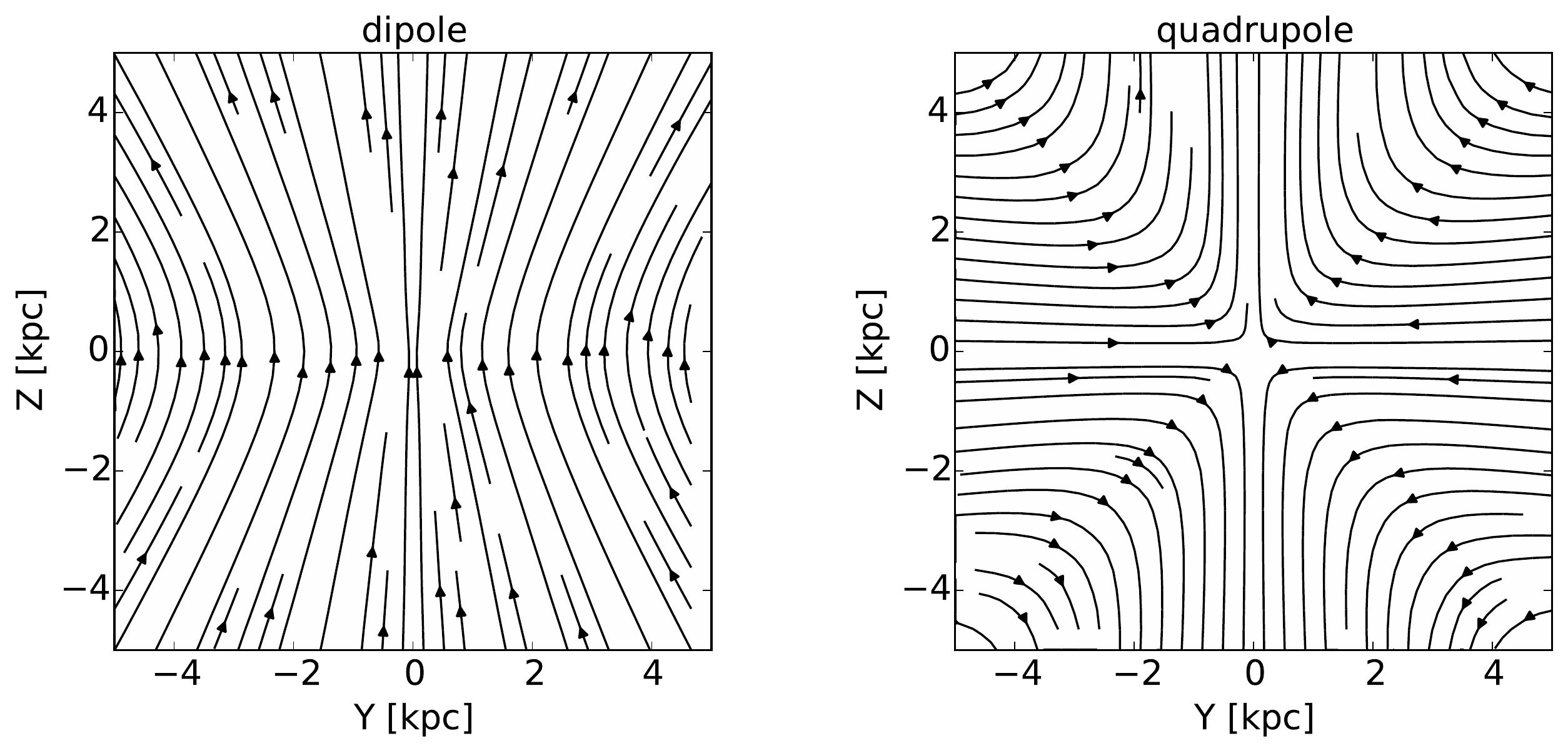}
\caption{Streamlines of the magnetic field for the two topologies used in the initial conditions. The left panel shows the dipole with mid plane symmetry of the vertical component and mid plane antisymmetry of the radial component, while the right panel shows the quadrupole field topology with opposite symmetries.}
\label{fig:initial-magnetic-field}
\end{figure*}

A fundamental ingredient in any MHD simulation is the adopted initial magnetic field configuration.  
The simplest possible choice would be a constant field parallel to one direction, e.g. a vertical uniform field
\begin{equation}\vect B\subscript{initial} = \begin{pmatrix}0\\0\\B_{0}\end{pmatrix}.\end{equation}
However, we argue here that this choice is not appropriate for initially concentrated mass distributions in general and for cosmological halos in particular. This simple
choice results indeed in a completely uniform magnetic energy distribution. As collapse proceeds, because of the initially peaked density distribution, less and less mass
is added to the central galaxy, especially at late time.  Magnetic energy, however, is still being accreted efficiently, even at late time, and artificially added to the central object.  
We believe it is more appropriate, in the context of ideal MHD (frozen-in magnetic flux), to consider that the initial magnetic energy follows closely the initial density distribution, 
in which case the magnitude of the field would scale roughly as
\begin{equation}
\left\Vert\vect B \right\Vert \propto \rho^{2/3}
\end{equation}
This requirement translates into a more complex field topology, and we need to work harder to initialise the field, compared to the uniform, vertical field case. 
Another important property of galactic dynamos is the field parity with respect to the system's mid-plane. To explore possible effects related to the direction of the
vector field, with odd or even parity across the mid plane, we consider initially two typical topologies: dipole-like or quadrupole-like. 
The dipole-like field (we use the acronym D) is defined using the following vector potential
\begin{equation}
\vect A\subscript{D} = B_{0}\left[\frac{\rho(r,z)}{\rho_{0}}\right]^{2/3}  r \vect e \subscript{\phi}
\end{equation}
where $\rho(r,z)$ is the initial gas density given by the NFW profile and $\vect e \subscript{\phi}$ is the unit vector along the toroidal direction. Note that here, coordinates are given in a cylindrical coordinate system centered around the halo center and aligned with the rotation axis. The corresponding magnetic field
has a  vertical component which is symmetric with respect to the mid plane, while its radial component is antisymmetric. The quadrupole field (we use the acronym Q) is defined using the vector potential
\begin{equation}
\vect A\subscript{Q} = B_{0}\left[\frac{\rho(r,z)}{\rho_{0}}\right]^{2/3} z \, \vect e \subscript{\phi}
\end{equation}
and has the reversed symmetries. Its vertical component is antisymmetric and its radial component is symmetric with respect to the mid plane. \autoref{fig:initial-magnetic-field} illustrates the shape of the two vector fields.
Note that in both cases, the initial field strength follows a profile peaked around the center with magnitude roughly proportional to  $\rho^{2/3}$, as expected. 
The magnetic field is then initialised on each cell face of the AMR grid, using a finite-difference approximation of the curl:
\begin{equation}\vect B\subscript{initial} = \nabla \times \vect A\end{equation}
This ensures that the divergence of the magnetic field is initially exactly zero (to machine precision), and, thanks to the Constrained Transport method, remains zero during
the entire simulation.

In the present paper, we would like to explore the purely kinematic regime of the magnetic field evolution. This corresponds to very small values of the magnetic field, for which there is no back reaction on the flow (the Lorentz force can be ignored). We therefore only solve the induction equation, which is linear with respect to the magnetic field. The exact value of the parameter $B_0$ is therefore irrelevant, and we will always quote magnetic field intensity as a function of the initial intensity, or as a function of the average intensity. We will study the saturation regime, and how the field reaches equipartition with the thermal and kinetic energies of the gas in a companion paper (hereafter Paper II). Close to saturation and equipartition, the exact value of the field matters a lot, and in this case, the initial intensity plays a very important role. Here, however, only the initial spatial distribution and the initial topology of the field are important, but not its overall initial normalisation. 

\subsection{Summary of additional physics parameters}

The feedback mechanism, whose details were explained in the previous section, can be switched on and off and in case of radiative feedback its effective strength can be controlled by changing the surrounding dust opacity parameter $\kappa$. The dwarf halo simulations were run without any radiative feedback. The only option was to have supernova feedback (simulations dubbed 'SN') or no feedback at all. In the Milky Way case, however, adding to those two options we also tested two more set-ups with radiation feedback. The values tested were one medium scale dust opacity value of $\kappa = 5 \unit{cm^{2}/g}$ ('K-5') for 70 K dust and a rather opaque gas with $\kappa = 20 \unit{cm^{2}/g}$ ('K-20') corresponding to a dust temperature of 140 K (cite Semenov et al. 2003). The feedback parameters for the dwarf as well as the Milky Way simulations are given in \autoref{haloic} and \autoref{feedbackic}. Both configurations have the same star formation $\epsilon_{*}=1\%$ and supernova feedback $\eta_{SN} = 10\%$ efficiencies. The temperature floor used to prevent the gas from fragmenting below our resolution limit is given by
\begin{equation}
T_{min} = T_{*} \left(\frac{n}{n_{*}} \right)^{2/3}
\end{equation}
where the critical temperature T$_*$ is a cool 100 K for the dwarf halo and warm 2000 K in the Milky Way case.

\begin{table}
\centering
\caption{Halo initial conditions parameters (see text for details).}
\label{haloic}
\begin{tabular}{ c | c | c | c }
\hline
\hline
 parameter & Dwarf & Milky-Way & units \\
\hline
$R_{200}$ & 50 & 230 & kpc \\
$V_{200}$ & 35 & 160 & km/s \\
$M_{200}$ & $1.4 \times 10^{10}$ & $1.3 \times 10^{12}$ & M$_{\odot}$ \\
$\Delta$x & 18 & 84 & pc \\
$m_{\rm res}$ & $\changed{1.5} \times 10^{3}$ & $\changed{1.5} \times 10^{5}$ & M$_{\odot}$ \\
$m_{*}$ & $2.0 \times 10^{3}$ & $\changed{5.9} \times 10^{4}$  & M$_{\odot}$ \\
T$_{*}$ & 100 & 2000 & K \\
n$_{*}$ & 14 & 4 & H/cc \\
$\epsilon_{*}$ & 1 & 1 & \% \\
$\eta_{\rm SN}$ & 10 & 10 & \% \\
Z$_{\rm ini}$ & 0.05 & 0.05 &  Z$_{\odot}$\\
met. yield & 10 & 10 & \% \\
\hline
\end{tabular}
\end{table}

\begin{table}
\centering
\caption{Initial magnetic field topology and feedback parameters (see text for details).}
\label{feedbackic}
\begin{tabular}{ l | c | c | c | c }
\hline
\hline
Name & Topology & SN Feedback & Opacity $\kappa$ [cm$^{2}$/g] \\
\hline
DW-D & Dipole & No & 0 \\
DW-D SN & Dipole & Yes & 0 \\
DW-Q & Quadrupole & No & 0 \\
DW-Q SN & Quadrupole & Yes & 0 \\
\hline
MW & Dipole & No & 0 \\
MW SN & Dipole & Yes & 0 \\
MW K-5 & Dipole & Yes & 5 \\
MW K-20 & Dipole & Yes & 20 \\
\hline
\end{tabular}
\end{table}

\section{Magnetic fields amplification through feedback processes}
\label{chap:results}
We will now present the results of our halo simulations, where we studied the influence of various stellar feedback parameters. This section is organised as follows: First, we present our dwarf galaxy simulations, without feedback, then using supernovae feedback. Second, we present the Milky-Way-sized galaxy simulations. For the latter case, supernovae feedback does not differ strongly from the no-feedback case, although it introduces slightly more turbulence in the gas. Radiation feedback makes however a big difference, and we explore two different dust opacities, resulting into two different scenarios for the galactic outflows.

\subsection{Dwarf galaxy}

\begin{figure*}
\centering
\includegraphics[width=\columnwidth]{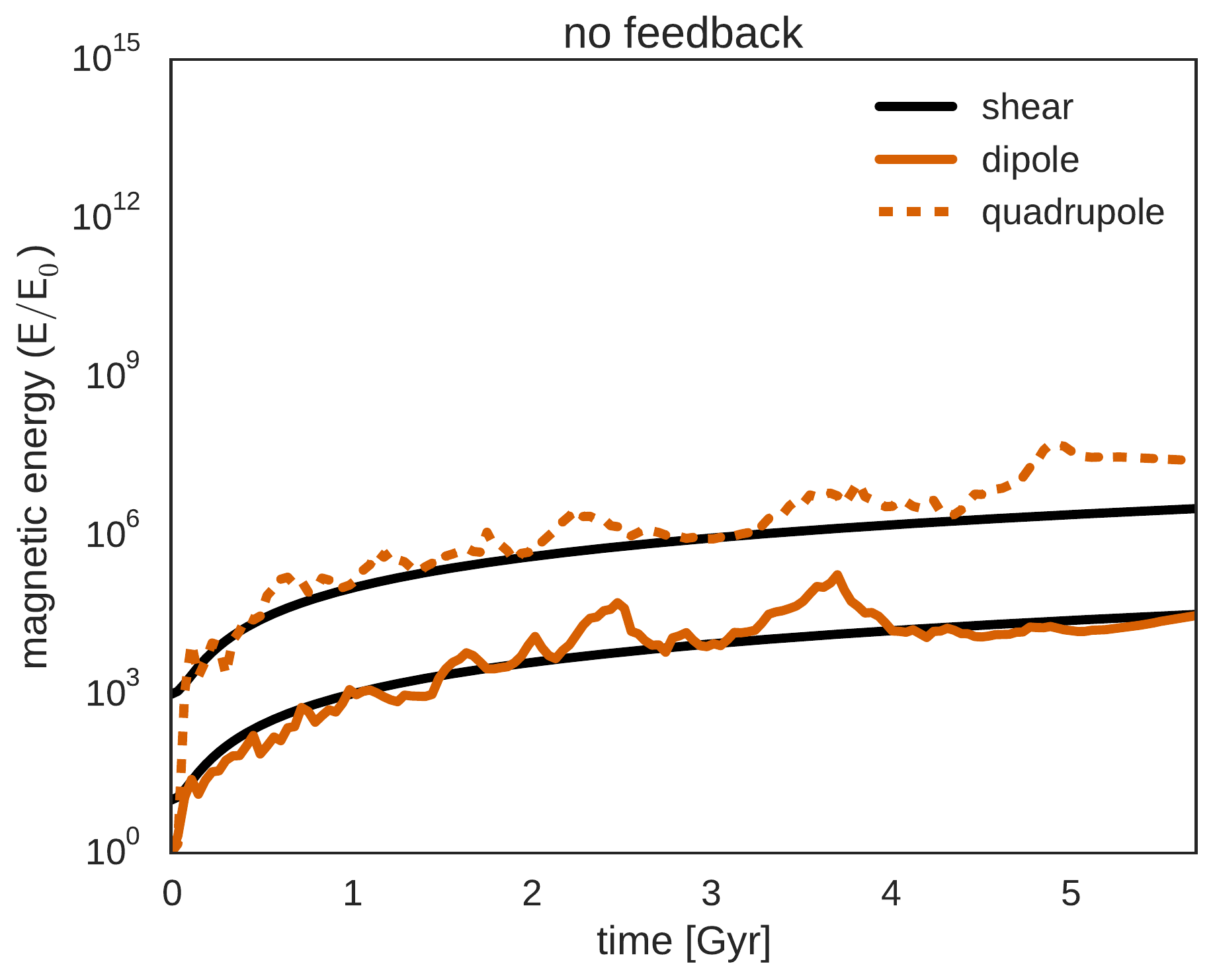}
\includegraphics[width=\columnwidth]{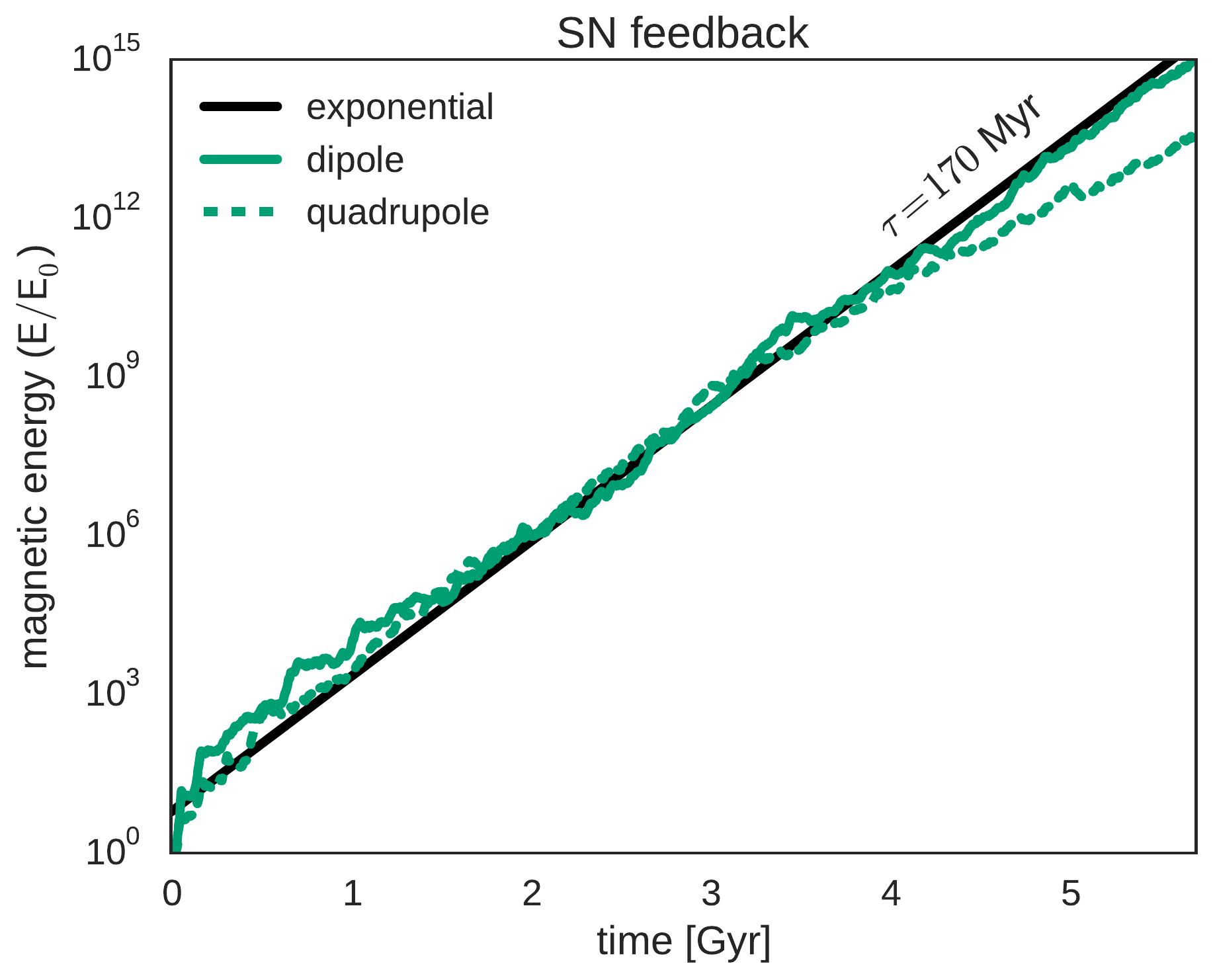}
\caption{\changed{Magnetic energy evolution in the dwarf galaxy simulations without feedback (left) and with supernova feedback (right), with dipole (solid lines) and quadrupole (dashed lines) initial conditions. Values are normalised to the initial total magnetic energy in the box. The black curves (left) illustrate the shearing amplification E$_{\mathsf{S}}$ with compression values of $\mathsf{E_{C}}=10 \mathsf{E_{0}}$ (dipole) and $\mathsf{E_{C}}=10^3 \mathsf{E_{0}}$ (quadrupole) and a shearing rate of $\mathsf{S} = (100 \mathsf{Myr})^{-1}$, while the black straight line (right) marks an exponential growth $\mathsf{exp}(t/\tau)$ at a rate of $\tau = 170 \mathsf{Myr}$ for comparison. Without feedback, magnetic energy rises a few orders of magnitude during the initial collapse but is from then only slightly amplified by shearing over the entire simulation time. With feedback, we observe a fast exponential growth of magnetic energy over many orders of magnitude.}}
\label{fig:dwarfenergy}
\end{figure*}

All our simulations begin in a similar way, which is the classical scenario for these cooling halo set-up. The gas, although initially in strict hydrostatic equilibrium, loses thermal energy through radiative cooling. It thus collapses and a centrifugally supported disk forms from the inside out, thanks to the initial angular momentum profile. In the dwarf galaxy case, the disc is relatively thick at first: Atomic cooling sets a natural temperature floor around $10^4$~K. Low temperature radiative processes (here mostly fine-structure cooling of metals) cools the gas further, leading to the formation of a thin disc which quickly fragments into dense gas clumps. The gas density in these clumps reaches the star formation density threshold and the first stars form.

\subsubsection{No feedback case}

In absence of feedback, the disc remains very thin, and the gas clumps are long-lived. Although our star formation efficiency was set very low (one percent), most of the gas inside the dense clumps is converted into stars, after a few disc orbital time. The resulting galaxy is very efficient at transforming most of its baryons into stars, which is at odd with observed dwarf galaxies in the universe. Moreover, the resulting circular velocity profile is strongly peaked towards the centre, although dwarf galaxies circular velocity profiles are usually declining towards the centre. 

The magnetic energy evolution can be seen in \autoref{fig:dwarfenergy}. Our new simulations confirm the earlier finding of \cite{2009ApJ...696...96W} and \cite{2010A&A...523A..72D}: During the early magnetic field amplification due to the collapse of the gas in the first few Myr, the magnetic energy increases as $\Delta^{4/3}$, where $\Delta$ is the ratio of the gas density after and before the collapse. Note that the magnetic field topology matters a lot in this early evolution. Without feedback, the dipole configuration leads to magnetic reconnection in the mid plane, as expected from the antisymmetry in the poloidal field. In the quadrupole field configuration, because of the symmetry of the poloidal field, the magnetic energy is not affected by field cancellation effects.

After the collapse, one can see in \autoref{fig:dwarfenergy} that the magnetic energy still grows, but much more slowly. This can be explained from field lines being twisted by the differential rotation. In this almost perfectly axisymmetric geometry, one can indeed approximate the induction equation as \citep[see e.g.][]{2010A&A...523A..72D}
\begin{equation}
\label{eqn:simpleindeqn}
\partial_t B_{r} \simeq 0
~~~\rm{and}~~ 
\partial_t B_{\theta} \simeq rB_r \partial_r \Omega \rm{.}
\end{equation}
The toroidal field grows therefore only linearly with time, while the poloidal field (mostly radial) remains constant. \changed{This results in quadratic time relation of the magnetic energy
\begin{equation}
E_{\mathsf{S}} = E_{\mathsf{C}} \cdot \left(1 + \left(S\cdot t\right)^2\right)
\end{equation}
with the magnetic energy after collapse E$_{\mathsf{C}}$, which depends on the field topology, and the shearing rate $S = r \partial_r \Omega$. We illustrate the contribution of this model for shearing amplification to $\mathsf{E_{mag}}$ for the no feedback simulations in \autoref{fig:dwarfenergy} (left).}

On the other hand, one can see directly from \autoref{eqn:simpleindeqn} that
a fast, exponential amplification of the field can be obtained only if the radial component grows as fast as the tangential component. For two dimensional, axisymmetric flows like our smooth rotating disk, this cannot be the case, according to the famous Zel'dovich and Cowling {\it anti-dynamo theorems}  \citep{Charbonneau:2012ts}. A further inspection of  \autoref{fig:dwarfenergy} reveals several spikes in the magnetic energy evolution. These are due to collapsing, rotating gas clumps, which trigger short episodes of field amplification. As anticipated by \cite{2009ApJ...696...96W}, these vortex modes do indeed amplify the field locally, but as soon as the clumps dissolve in the large-scale rotating flow, so does their magnetic energy.

\begin{figure*}
\centering
\includegraphics[width=\columnwidth]{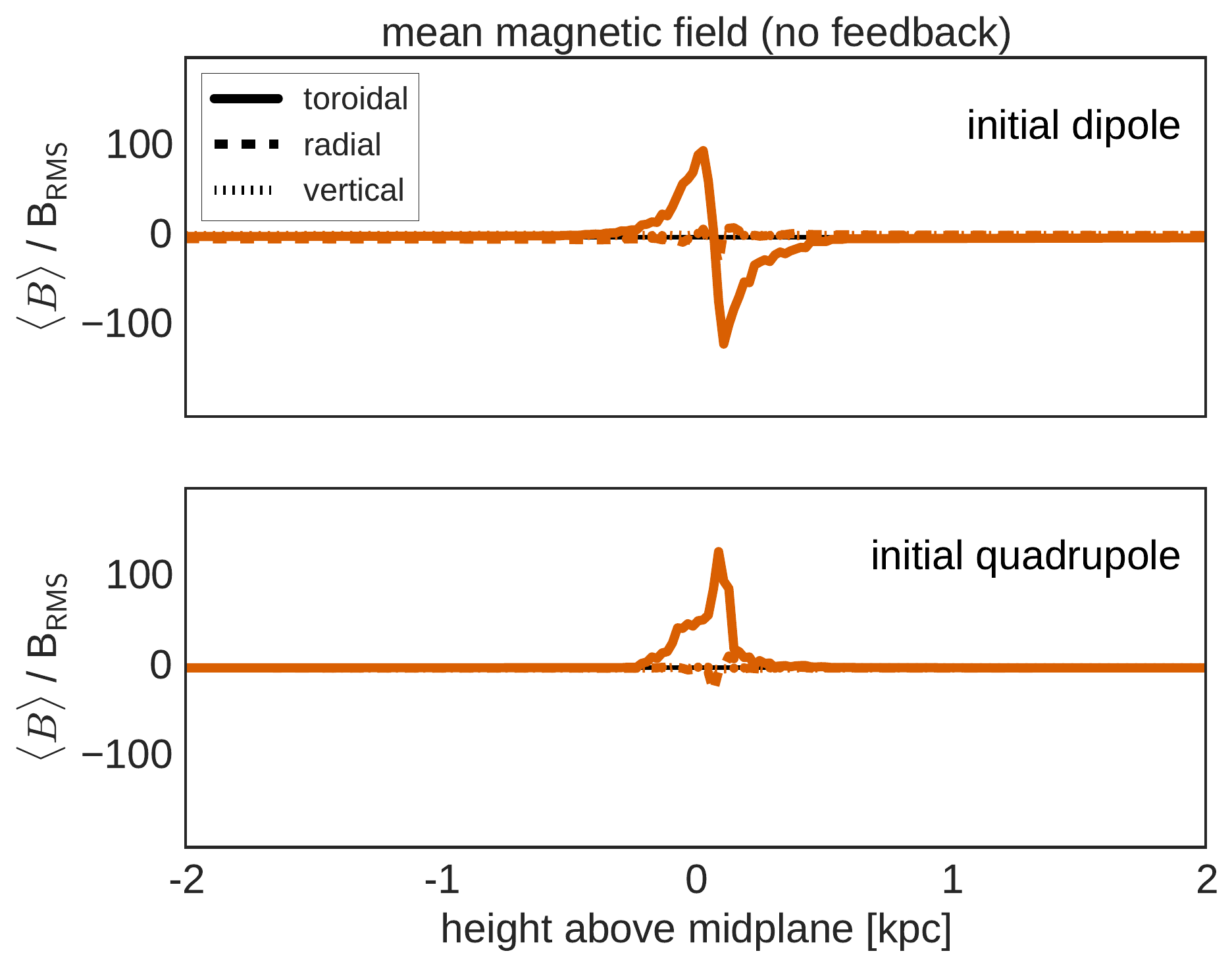}
\includegraphics[width=\columnwidth]{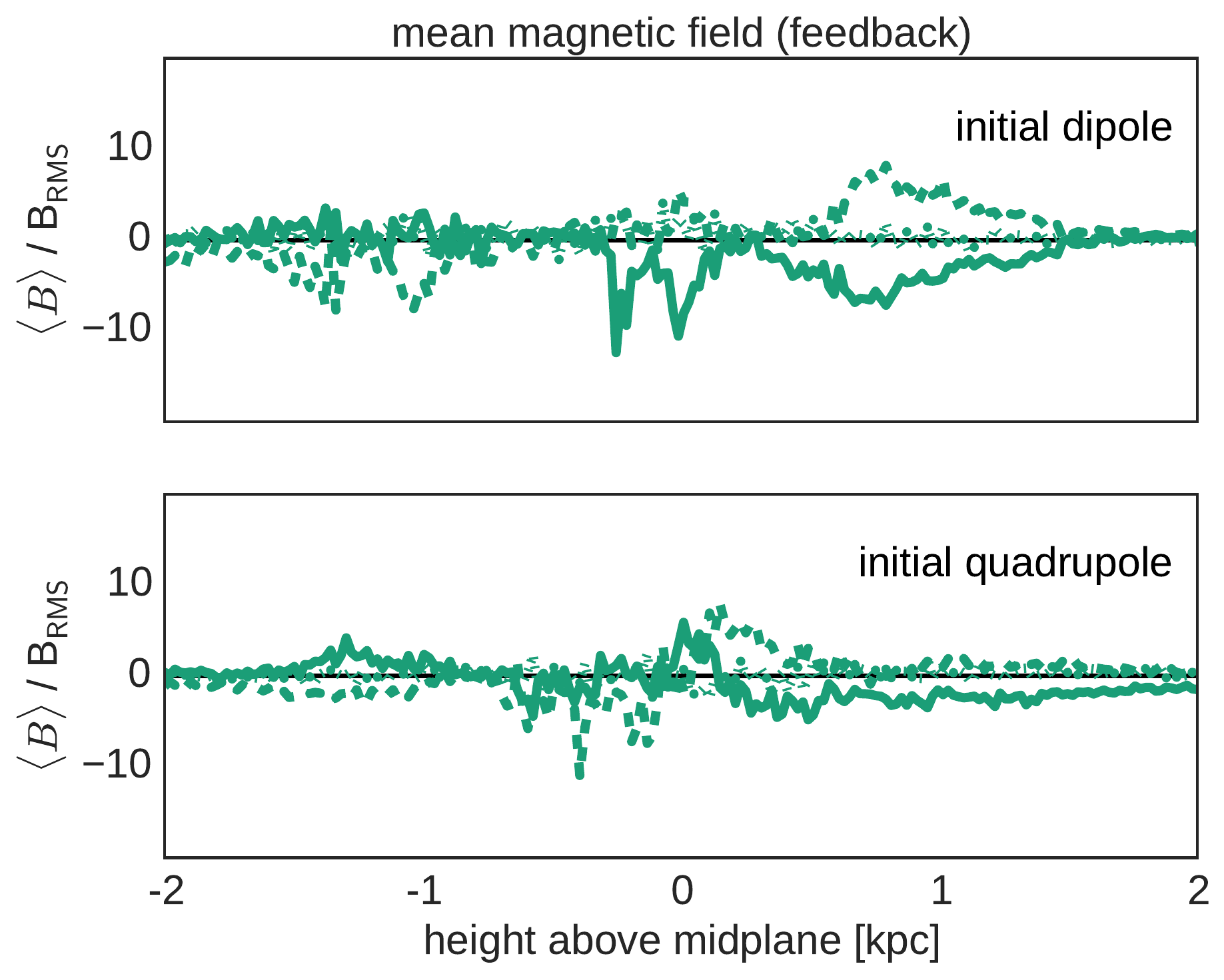}
\caption{Magnetic field components after 1.9 Gyr, as a function of the vertical height relative to the galactic mid plane, normalised to $\upright{B}\subscript{RMS}$, the root mean square (rms) value of the field amplitude. Each component has been computed for each z bin as the volume-weighted average value inside each slice (the bin size is 10 pc). Without feedback, the initial mid plane symmetry of the radial component is imprinted on the toroidal field, whose peak value is rather high (in units of the rms field). With feedback, all 3 field components are equally strong and their typical values comparable in strength to the rms.}
\label{fig:dwnofbk-midplane}
\end{figure*}

These clumps also trigger three-dimensional turbulence in the gas, thanks to clump-clump interactions \citep{2009MNRAS.392..294A}. This could in principle increase the magnitude of the radial component of the magnetic field, but the induced effects remain too weak to affect the large-scale dynamo. \autoref{fig:velprofdw} shows the velocity dispersion of the dwarf galaxy in the no-feedback case: It barely reaches 10\% of the tangential velocity. As a consequence, the corresponding magnetic field remains mostly toroidal, as shown in \autoref{fig:dwnofbk-midplane}. One can also clearly see in this Figure that the initial field parity (odd for the dipole and even for the quadrupole) has been conserved during the collapse and the subsequent shear amplification, providing a direct dependence of the final field parity on the initial halo field topology.

\begin{figure*}
\centering
\includegraphics[width=0.667\linewidth]{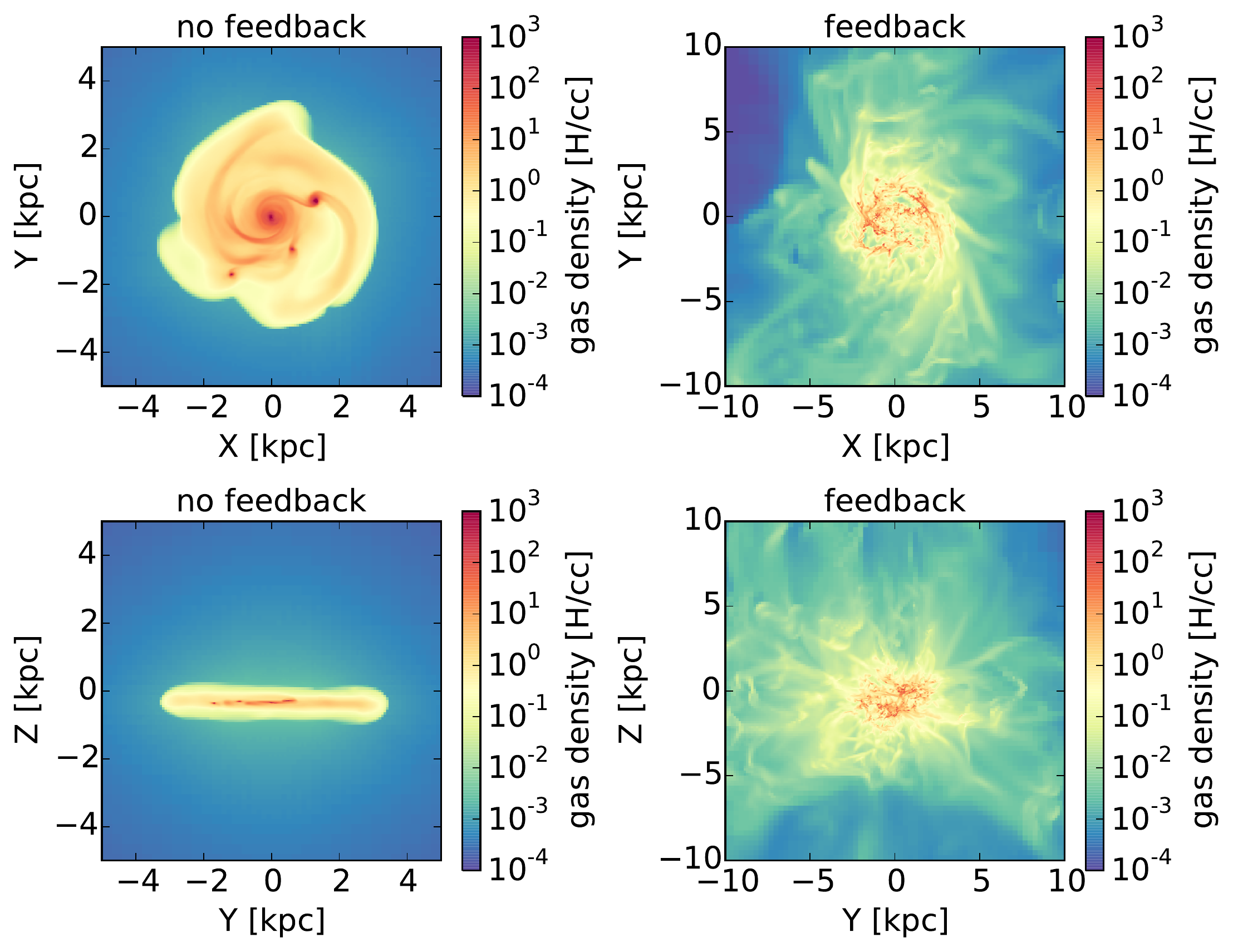}
\caption{Face-on (top row) and side-on (bottom row) of mass-weighted density projections in the dwarf galaxy simulation at 1.9 Gyr. When there is no feedback (left column), the gas builds up a thin and rotationally suported disk. With feedback (right column), we form a thick, turbulence-supported disk with strong outflows.}
\label{fig:dwmaps-rho}
\end{figure*}

\begin{figure*}
\centering
\includegraphics[width=0.667\linewidth]{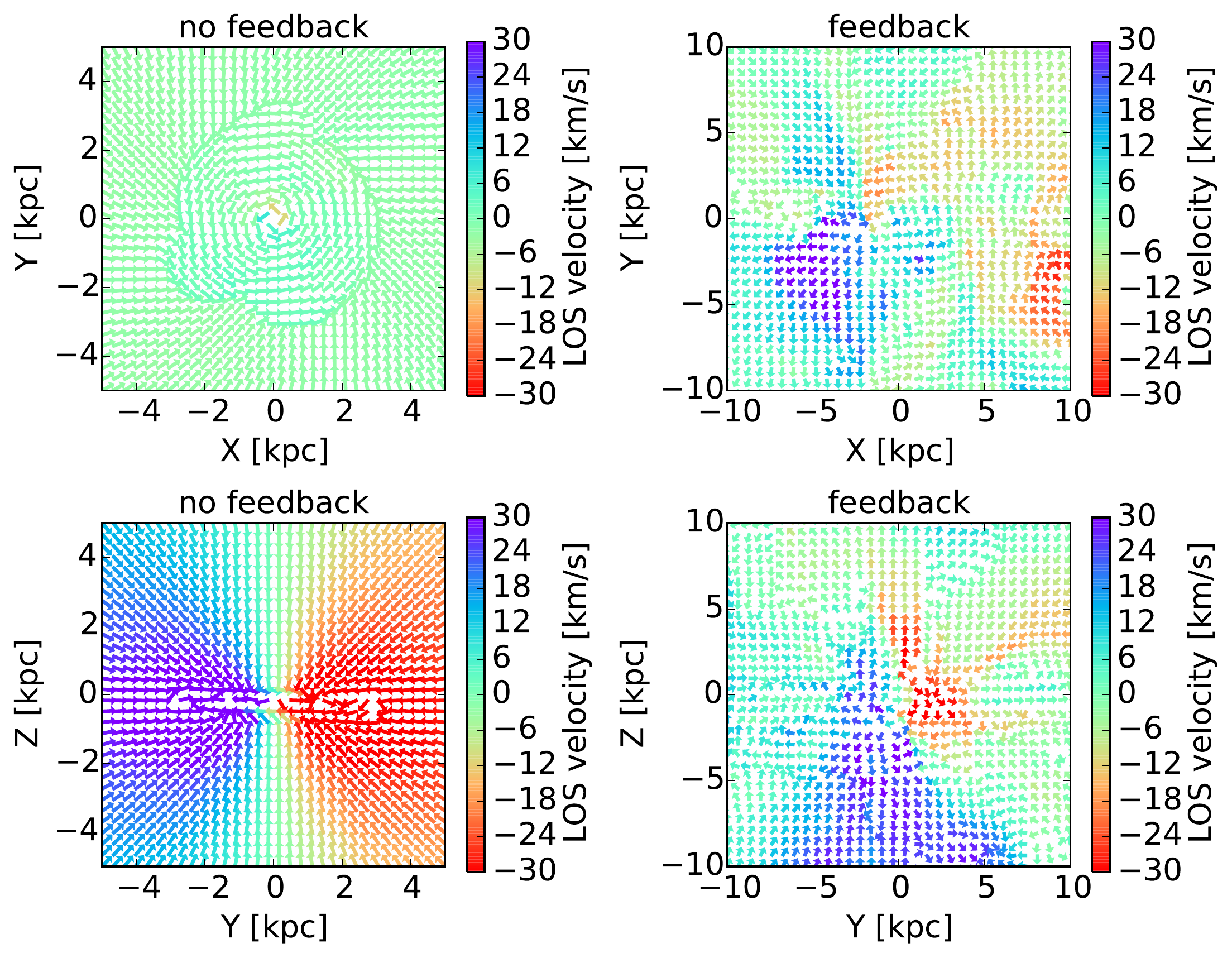}
\caption{Mass-weighted average velocity field maps corresponding to the density projections of \autoref{fig:dwmaps-rho}. Arrows show direction (but not strength) of the mean velocity field perpendicular to the line of sight. Colours indicate the strength and direction of the line-of-sight velocity component, where blue means approaching and red means receding from the observer. Without feedback (left), gas motion is dominated by global rotation (bottom) and the vertical velocity component is weak (up). Feedback (right) drives winds which can be seen as a strong vertical component (upper panel) and significant deviations from pure rotation (lower panel).}
\label{fig:dwmaps-vel}
\end{figure*}

\begin{figure*}
\centering
\includegraphics[width=0.667\linewidth]{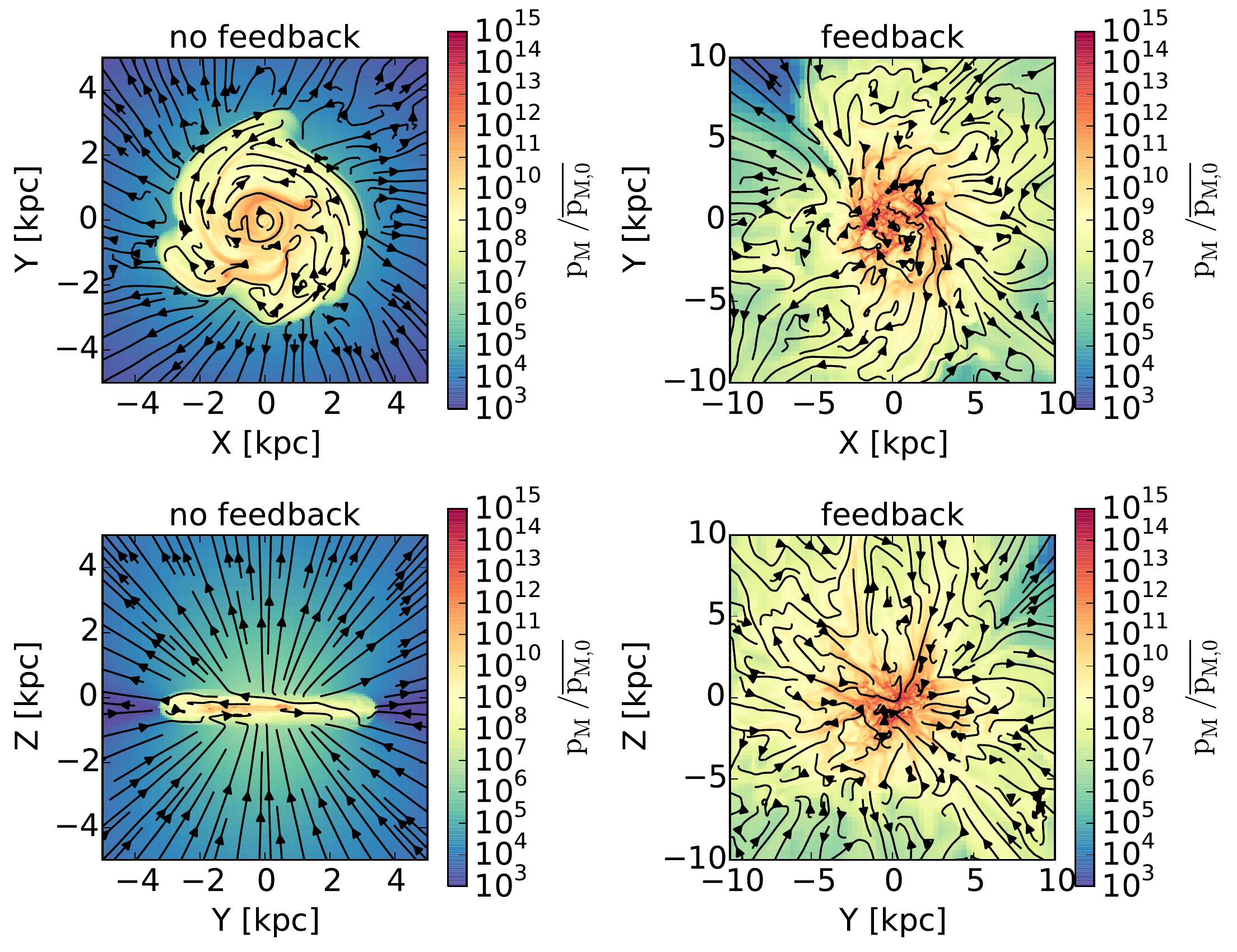}
\caption{Mass-weighted magnetic pressure maps of the dipole simulation at 1.9 Gyr, normalized to the initial average magnetic pressure. Overlaid in black are streamlines of the mean field perpendicular to the line of sight. Without feedback (left), the field is ordered and dominated by a large-scale dipole structure (bottom) and a dominant toroidal field in the disk (top). With feedback (right), the magnetic field is characterised by random turbulent motions, without preferred direction or large-scale pattern (both right panels).}
\label{fig:dwmaps-mag}
\end{figure*}

\subsubsection{Supernovae feedback case}

We now describe our results for the dwarf galaxy with supernovae feedback enabled. The evolution is drastically different, with violent outflows terminating quickly the life of the dense star-forming clouds. The resulting star formation rate is reduced by one order of magnitude, compared to the no-feedback case. As shown in \cite{2013MNRAS.429.3068T}, the galactic circular velocity is now in much better agreement with observed dwarf galaxies, exhibiting a kpc-sized core in the dark matter distribution. Star formation also proceeds in successive starbursts, leading to the ejection of massive quantities of gas into a galactic fountain. The gas falls back after a dynamical time, triggering a new star formation event. 

\begin{figure}
\centering
\includegraphics[width=\columnwidth]{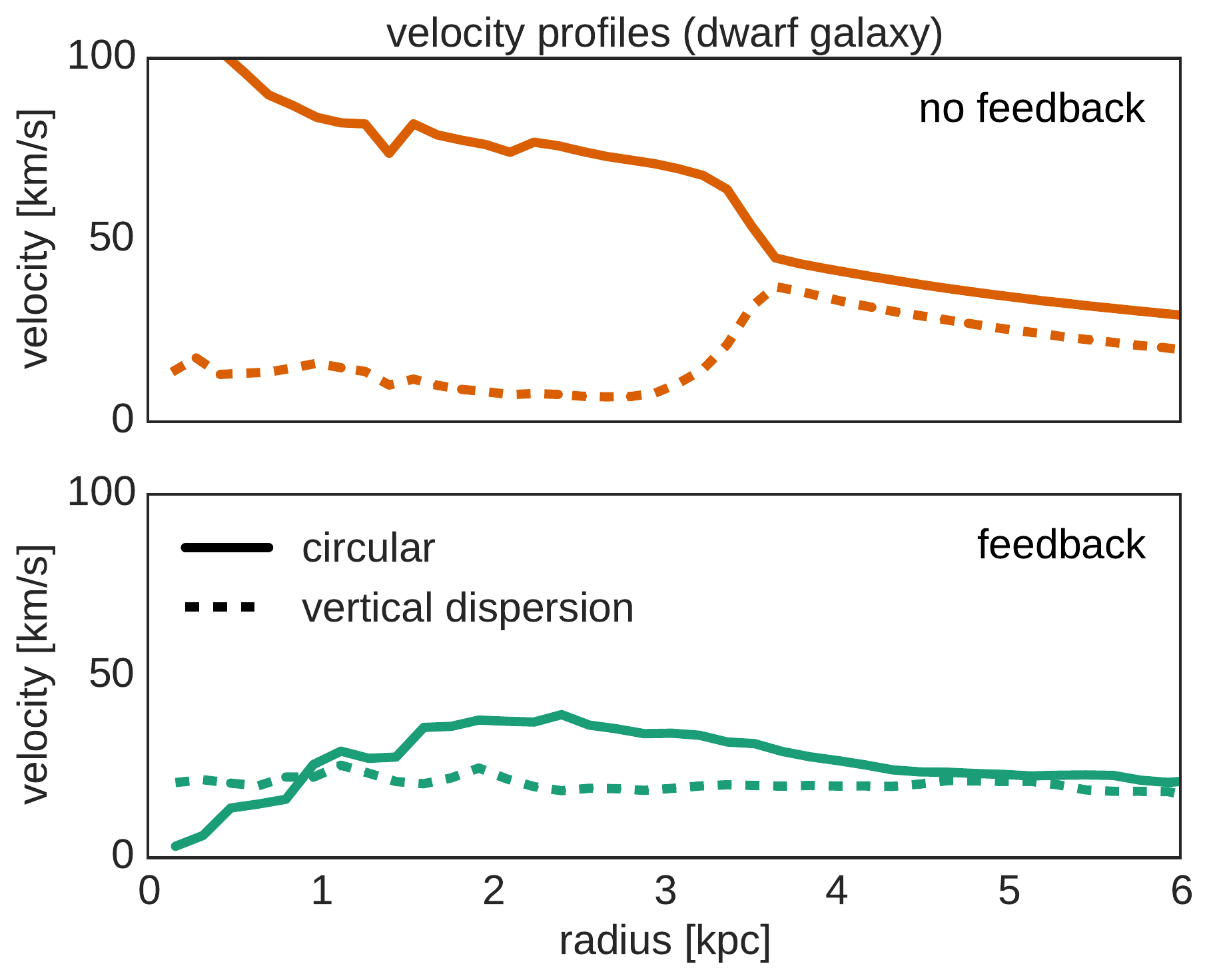}
\caption{Velocity profiles of circular velocity and vertical velocity dispersion in our dwarf galaxy simulations without (top) and with feedback (bottom) at 1.9 Gyr. The solid lines show the average rotational velocity $\overline{V_\theta}$  and the dashed lines shows the vertical velocity dispersion $\sigma_z$, both as a function of the cylindrical radius.}
\label{fig:velprofdw}
\end{figure}

The corresponding magnetic energy evolution can be seen in in \autoref{fig:dwarfenergy}. We observe, for both dipole and quadrupole initial conditions, a very fast, exponential growth with e-folding time of around 200~Myr. The measured growth rate is therefore quite fast, comparable to the rotation rate $\Gamma \simeq \Omega$. Note that the magnetic energy has been amplified by almost 18 orders of magnitude, which correspond to 9 orders of magnitude in the magnetic field itself. \autoref{fig:velprofdw} compares the rotational velocity $V$ to the vertical velocity dispersion $\sigma$. We have in the feedback case $V/\sigma \simeq 1$, a clear sign of strong turbulence. As a result, field lines are violently twisted in random directions, allowing the radial and vertical components of the field to reach a similar strength than the toroidal component. \autoref{fig:dwnofbk-midplane} illustrates this isotropy in the magnetic field. The field is now highly turbulent, with the largest fluctuation seen on the smallest scales (a few cell in size). 

\begin{figure}
\centering
\includegraphics[width=\columnwidth]{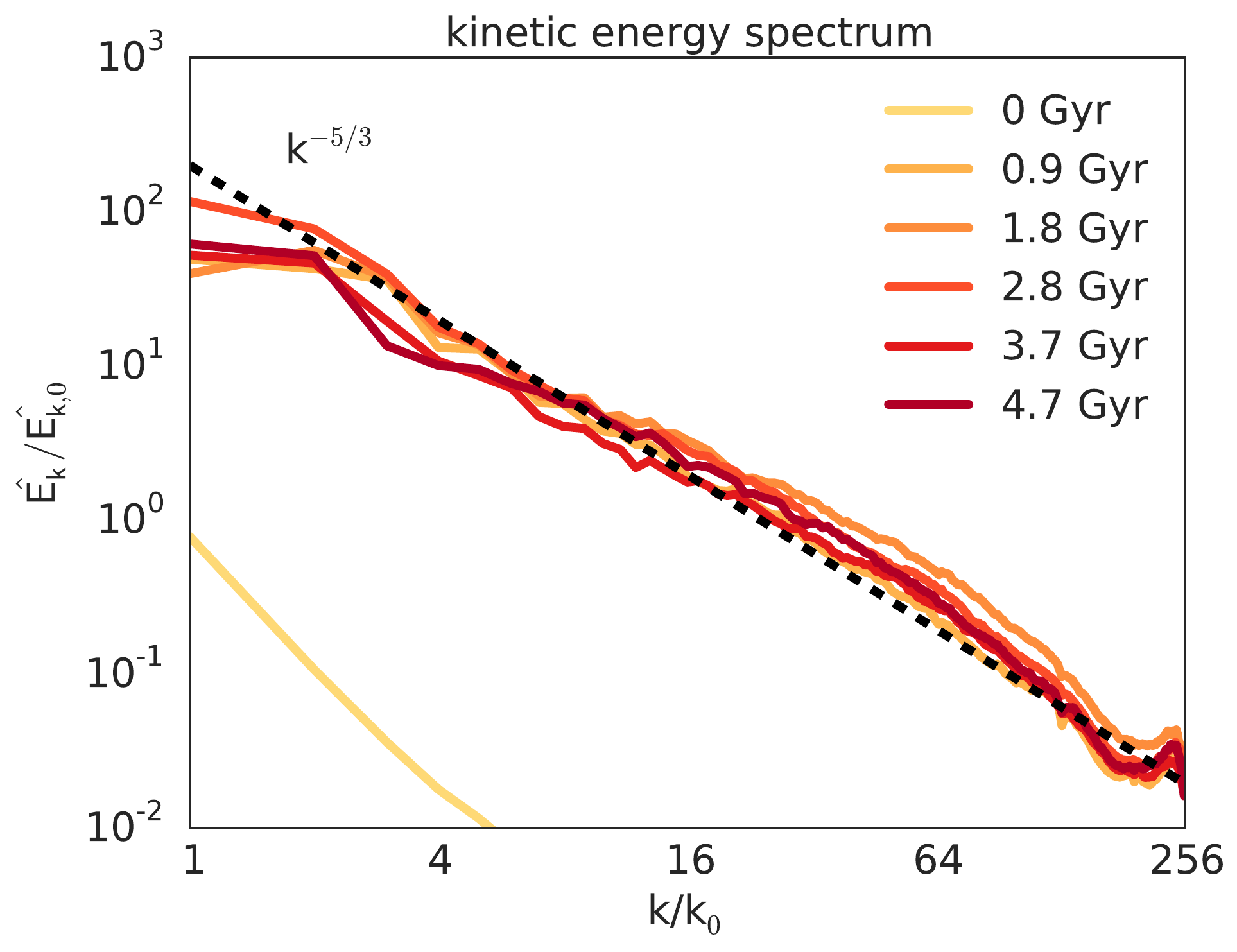}
\includegraphics[width=\columnwidth]{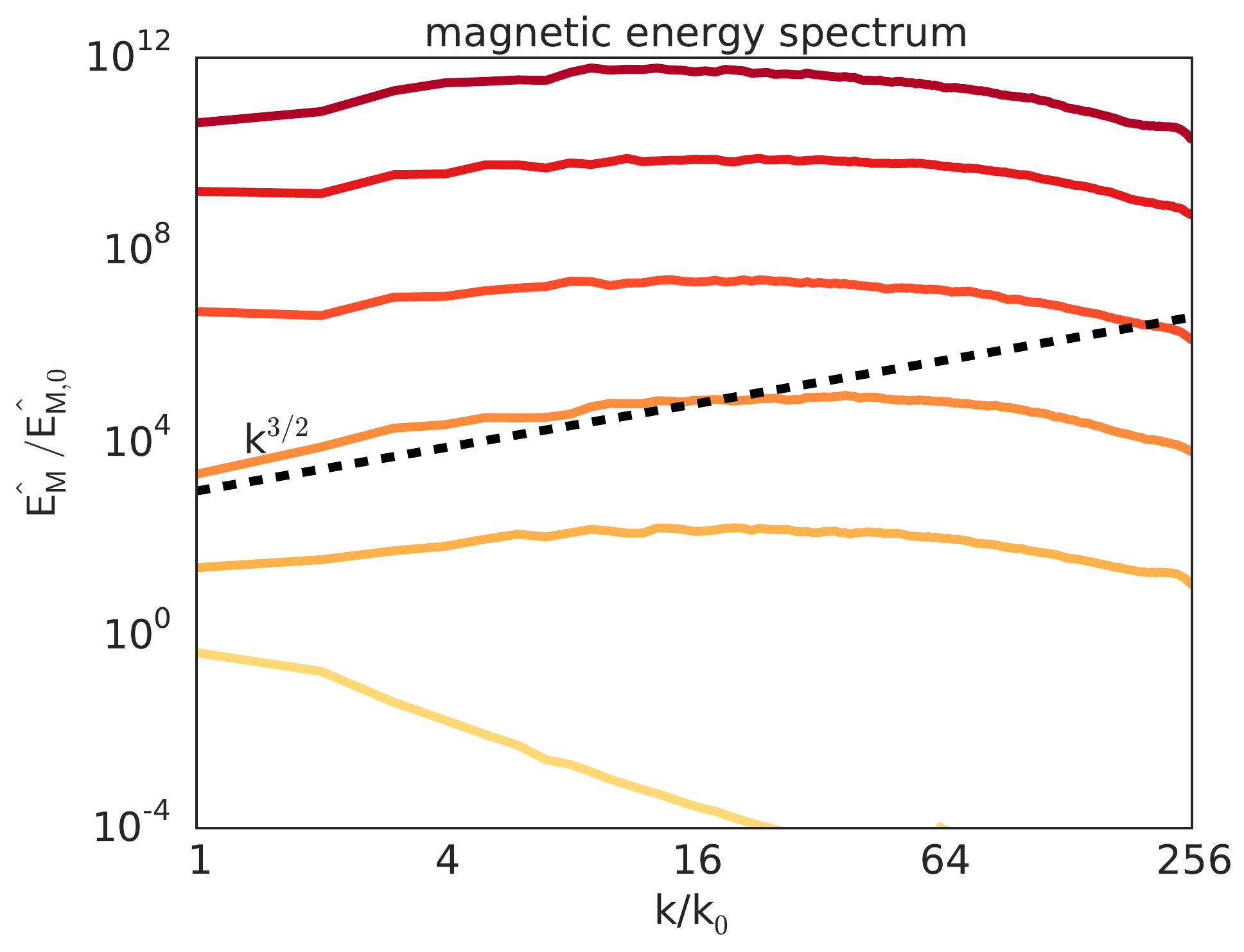}
\caption{Time evolution of the kinetic and magnetic energy spectrum in a central $512^3$ box and normalised to the total initial magnetic energy in the box. The simulation quickly develops a k$^{3/2}$, exponentially growing energy spectrum, typical of small-scale Kazantsev's dynamos, bottlenecked on scales of a few grid cells.}
\label{fig:spectra}
\end{figure}

To study further the interplay between the supernovae-driven turbulence and the growth of the field, we have computed the power spectra of both the gas kinetic energy and the gas magnetic energy in \autoref{fig:spectra}. For the former, we see the almost immediate onset of a power law power spectrum like the one predicted by Kolmogorov ($P \sim k^{-5/3}$), quite typical of supersonic turbulence. The kinetic energy power spectrum appears very stable throughout the evolution, maintained at this high level by supernovae explosions and rotational energy. Note that the injection scale of the kinetic energy is very large here: it is the size of the entire galaxy. To get an idea of the nature of the forcing of the turbulence in this feedback-dominated galaxy, we have plotted in \autoref{fig:dwmaps-vel} a rendering of the gas velocity field. It shows large-scale upward and downward motions, together with a clear overall rotation pattern. The largest scale at which kinetic energy is injected turns out the be roughly the halo scale radius $r_s$, for which marks the transition between the two power law regimes in the dark matter distribution (from $r^{-1}$ deep inside to $r^{-3}$ in the outskirts). In what follows, this radius will be identified to the kinetic energy injection scale, also noted $L$. Note that in our spectral analysis, the global rotation was not removed from the velocity field before computing its Fourier transform, because turbulence is clearly dominating. As an additional caveat, we also note that both the gas density and the magnetic field are far from being homogeneously distributed in the box where the spectrum is computed, so that the signals are only approximately isotropic and far from periodic.

The magnetic energy power spectrum, on the other hand, is plotted in \autoref{fig:spectra} (bottom). We see here again that its amplitude is exponentially growing, while its shape remains roughly the same, with $P \sim k^{3/2}$ on the large-scale end. The power spectrum reaches a maximum at a scale corresponding to 5 cells, then slowly declines as its gets to the Nyquist frequency of the grid. This is exactly what is predicted from Kazantsev's theory of turbulent dynamos \citep{1968JETP...26.1031K}, and confirmed in the forced-turbulence, periodic box MHD simulations of \cite{Haugen:2004fs}. In the present dwarf galaxy simulations, we also obtain a small-scale magnetic dynamo, for which the forcing scale would be the size of the entire galaxy $L \simeq 10$~kpc, and for which the magnetic dissipation scale would be set by the adopted numerical resolution. 

In the small-scale dynamo theory, a critical ingredient is the magnetic Reynolds number $R_m=V L / \eta$ which encode the magnitude of the small-scale magnetic dissipation. In our notation, the velocity dispersion at the forcing scale $L$ is $V$ and $\eta$ is the magnetic dissipation coefficient. As discussed in \cite{2005PhR...417....1B}, exponential growth of the field is obtained if $R_{m}> R_{\rm crit}$, where the critical magnetic Reynolds number was observed to be around 30-35. In our case, where no explicit magnetic dissipation has been included in the induction equation, this translates into a critical spatial resolution, beyond which we expect to see exponential amplification of the field. 

\begin{figure*}
\centering
\includegraphics[width=\columnwidth]{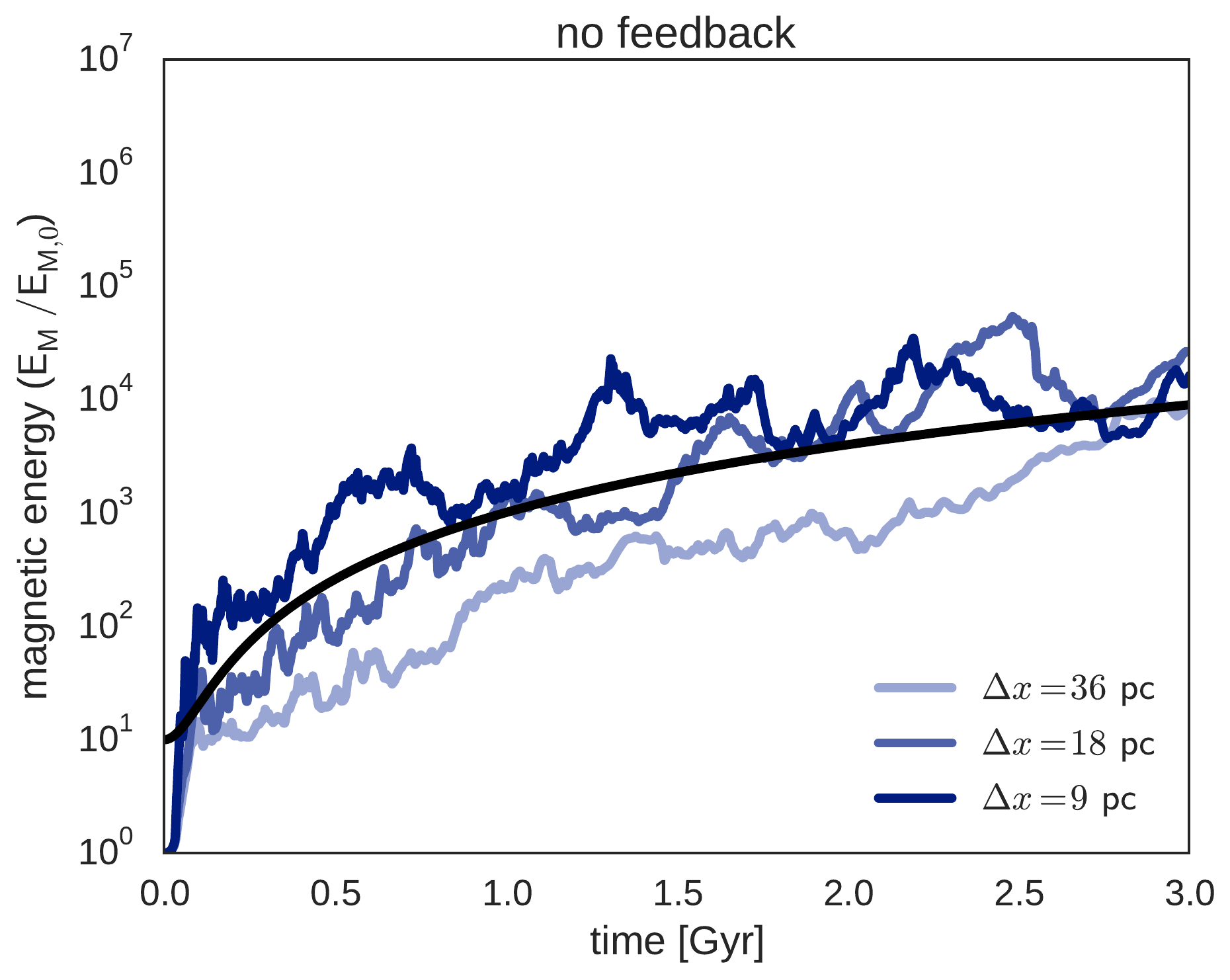}
\includegraphics[width=\columnwidth]{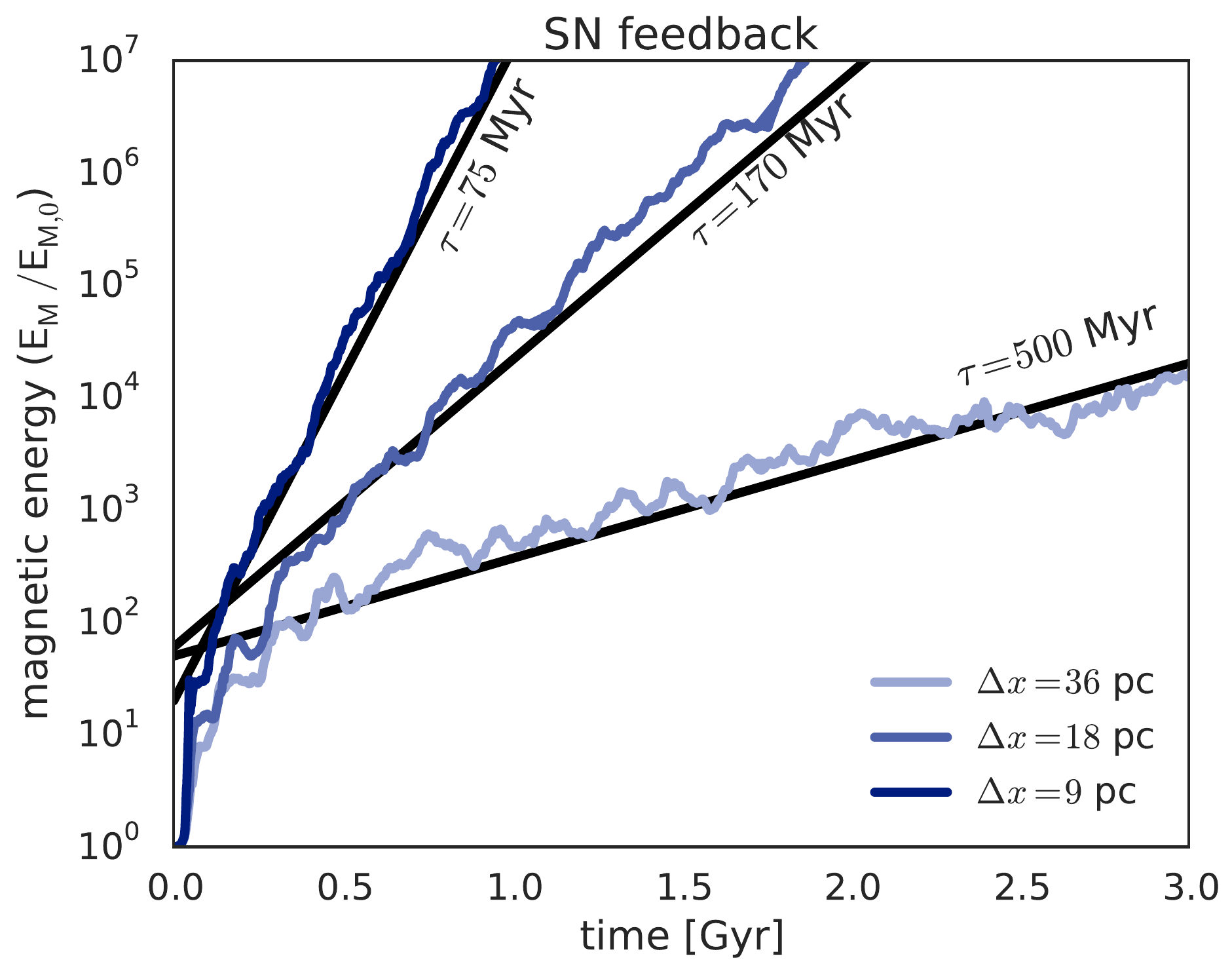}
\caption{\changed{Time evolution of total magnetic energy in the simulation box for the dwarf galaxy simulation with different maximum resolutions without feedback (left) and with supernova feedback (right) and normalised to their initial values. The initial magnetic field topology used in all cases is the dipole type. We explore the effect of half and double resolution as compared to the original resolution of $\Delta$x~=~18~pc. The black curve in the left plot illustrates the same shearing amplification as in \autoref{fig:dwarfenergy}. The black straight lines on the right-hand side show exponential growth \changed{exp($t/\tau$) with values of $\tau$ = 500, 150, and 70 Myr respectively}.}}
\label{fig:energy_resolutions}
\end{figure*}

\begin{figure*}
\centering
\includegraphics[width=\linewidth]{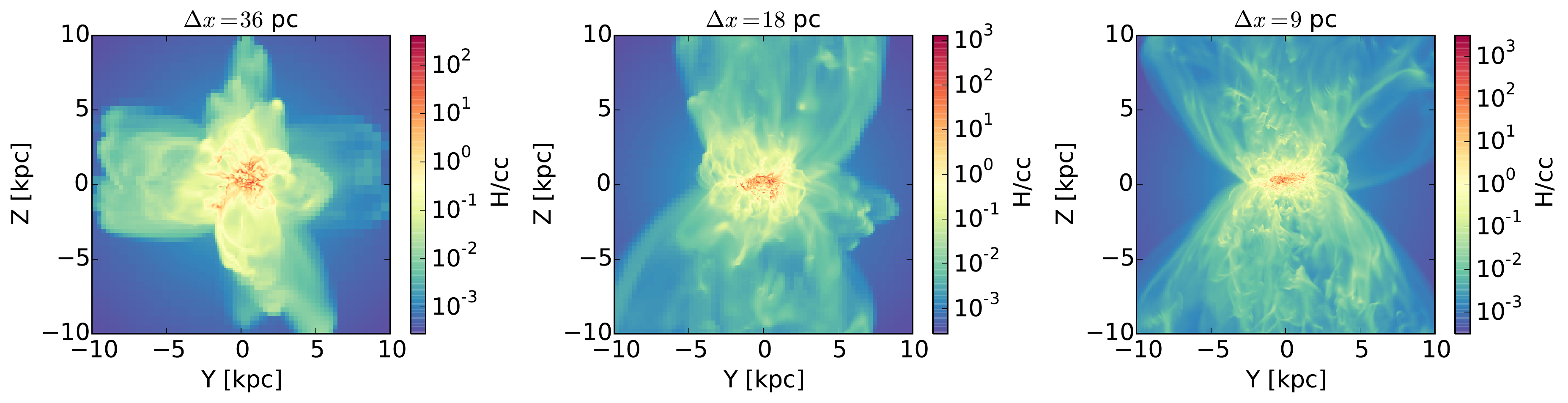}
\includegraphics[width=\linewidth]{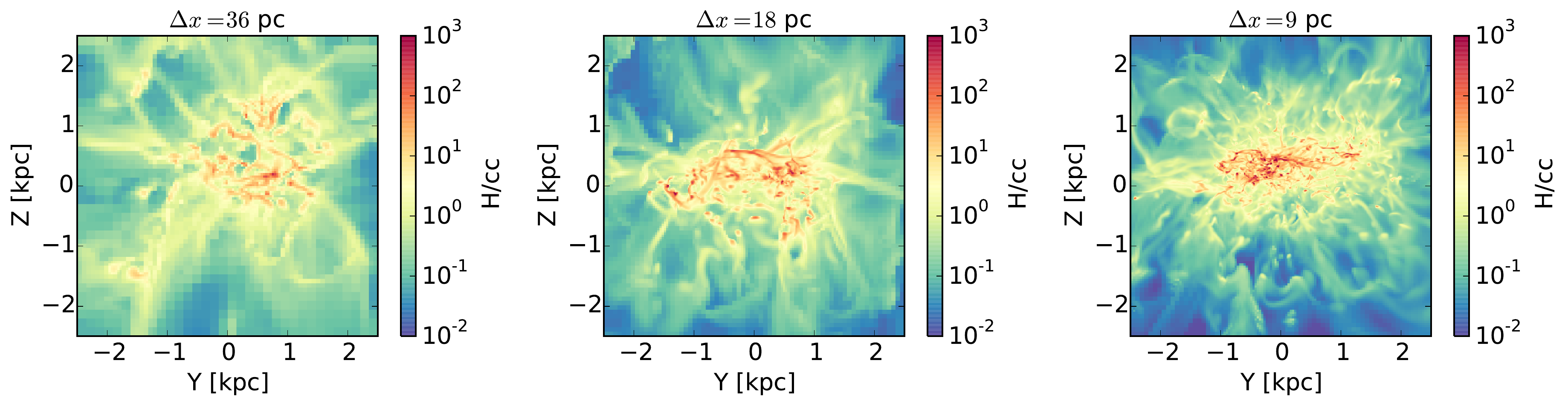}
\caption{\changed{Side-on views of the mass-weighted average density in the central region of the dwarf galaxy simulation with the different spatial resolutions as in \autoref{fig:energy_resolutions}, with a panel size of 20 kpc (top row) and 5 kpc (bottom row). The maximum resolution is increased from left to right, starting from $\Delta x = 36$ pc (left) to $\Delta x = 18$ pc (middle) $\Delta x = 9$ pc (right). Increasing the resolution allows one to reveal more substructures, finer details in the flow and stronger winds.}}
\label{fig:density_maps_resolution}
\end{figure*}

For this reason, we repeat the dwarf simulations at different resolutions to study their effect. \changed{We show in \autoref{fig:energy_resolutions} the time evolution of the total magnetic energy in our dwarf galaxy with 3 different maximum resolutions: $\Delta x=$36, 18 and 9~pc. We see a tendency for stronger amplification from compression with increasing resolution in all cases because the gas can build structures with higher density. Yet, without supernova feedback the subsequent shearing amplification does not show a significant impact of resolution in the resulting magnetic energy at the end. The feedback-driven dynamo, on the other hand, shows a strong growth rate dependence on the maximum resolution. In the low-resolution run, we obtain a rather slow amplification, with $\Gamma \simeq 0.4 \Omega$. This means we are close to or slightly better than the critical resolution for small-scale dynamo. For our fiducial resolution, we see a fast exponential growth with $\Gamma \simeq \Omega$. In the high resolution case, we observe an even faster growth with $\Gamma \simeq 2 \Omega$. In \autoref{fig:density_maps_resolution}, images of gas density projections are plotted to show how the gas structure is resolved at different resolutions. With increasing resolution, we see more dense substructures and importantly finer details in the flow and stronger winds.}

This behaviour can be explained nicely within the framework of small-scale dynamos, for which the growth rate is determine by the inverse of the eddy turn-over time at the dissipation scale $\ell$, namely $\Gamma \simeq v(\ell) / \ell \propto \ell^{-2/3}$ for Kolmogorov's turbulence. At higher resolution, the eddy turn-over time scale become shorter, so that the growth rate becomes larger. In the kinematic phase, well before the saturation phase, when the magnetic energy will reach equipartition with the kinetic energy, the growth rate of the small-scale dynamo is therefore determined by the smallest resolved scale of the turbulence, $\ell$, and propagates through an inverse cascade all the way up to the forcing scale $L$,
following the $k^{3/2}$ Kazantsev's power law. 

Since actual microscopic dissipation processes are occurring on very tiny scales, completely unresolved even by our highest resolution run, one expects the actual growth rate to be even higher than the already fast rates we have measured in our numerical experiments. It is therefore reasonable to assume that field saturation will be reached very quickly.  We however defer the study of the saturation phase to a follow-up, companion paper (hereafter Paper~II).

\subsection{Milky-Way-like galaxy}

We now describe our results for the Milky-Way-sized halo. In this case, the gas quickly cools down to $10^4$~K, which leads to the immediate formation of a very thin disk. 
Our spatial resolution in the Milky-Way case is limited to 84~pc, so we can't allow the gas to cool much below 3000~K, as summarised in \autoref{haloic}. Nevertheless, the disk is so massive that it also quickly fragments into massive clumps that actively form stars.

\subsubsection{No feedback and supernovae feedback cases}

In the no-feedback case, the galactic disk develops a massive central condensation of stars, which results in a central circular velocity close to 500~km/s. With supernovae feedback, however, we manage to reduce significantly the central bulge mass, with a maximum rotational velocity approaching only 250~km/s (see \autoref{fig:velprofmw}).
Note that even with supernovae feedback included, we do not reduce the overall star formation efficiency, and after several~Gyr, most of the baryons have been converted into stars.
This can be explained by the relatively low specific energy released by supernovae, making it very difficult for the gas to reach the escape velocity of the halo \citep[see for example][for a complete discussion]{2014MNRAS.444.2837R}. A small galactic fountain sets in, so that the gas remains mostly bound to a weakly turbulent disc, as can be seen on \autoref{fig:velprofmw} in the vertical velocity dispersion profile.

\autoref{fig:mwenergy} shows the magnetic energy evolution for these two cases, and they exhibit the same features as the no-feedback dwarf galaxy case: early magnetic field amplification due to gravitational collapse of the cooling halo gas, followed by a weak shear amplification, with some fluctuations associated with fragmenting-clumps-induced vortex modes. The corresponding magnetic field topology appears as very well organised on large scales, with a dominating toroidal component.  

\subsubsection{Radiative Feedback case}

\begin{figure*}
\centering
\includegraphics[width=\linewidth]{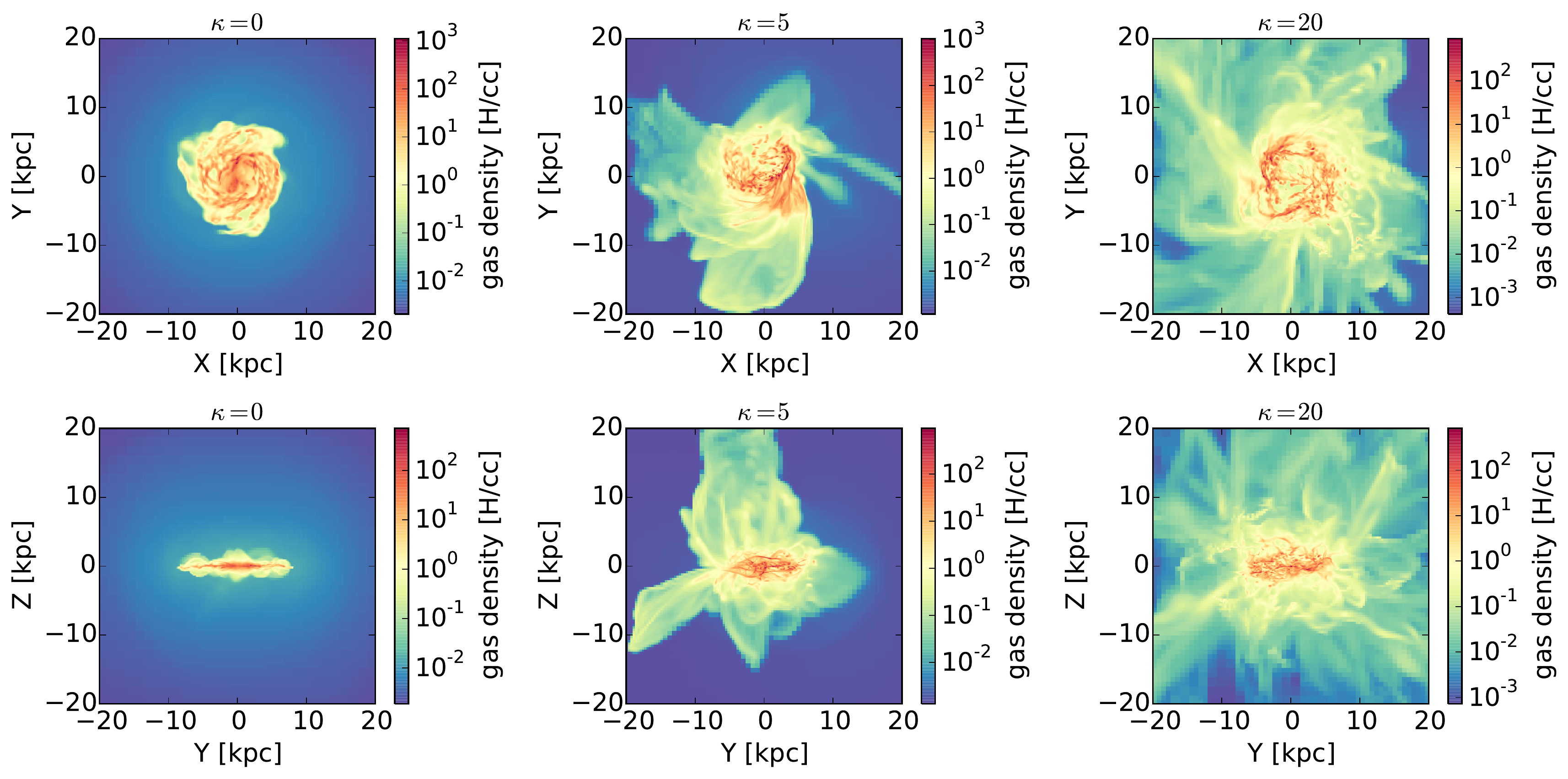}
\caption{Face-on (top row) and side-on (bottom row) views of mass-weighted average density in the Milky Way simulations at 2.3 Gyr. Radiative feedback efficiencies are low (left), medium (middle) and high (right). With increased feedback strength, the disk becomes thicker and winds stronger.}
\label{fig:mwmaps-rho}
\end{figure*}

\begin{figure*}
\centering
\includegraphics[width=\linewidth]{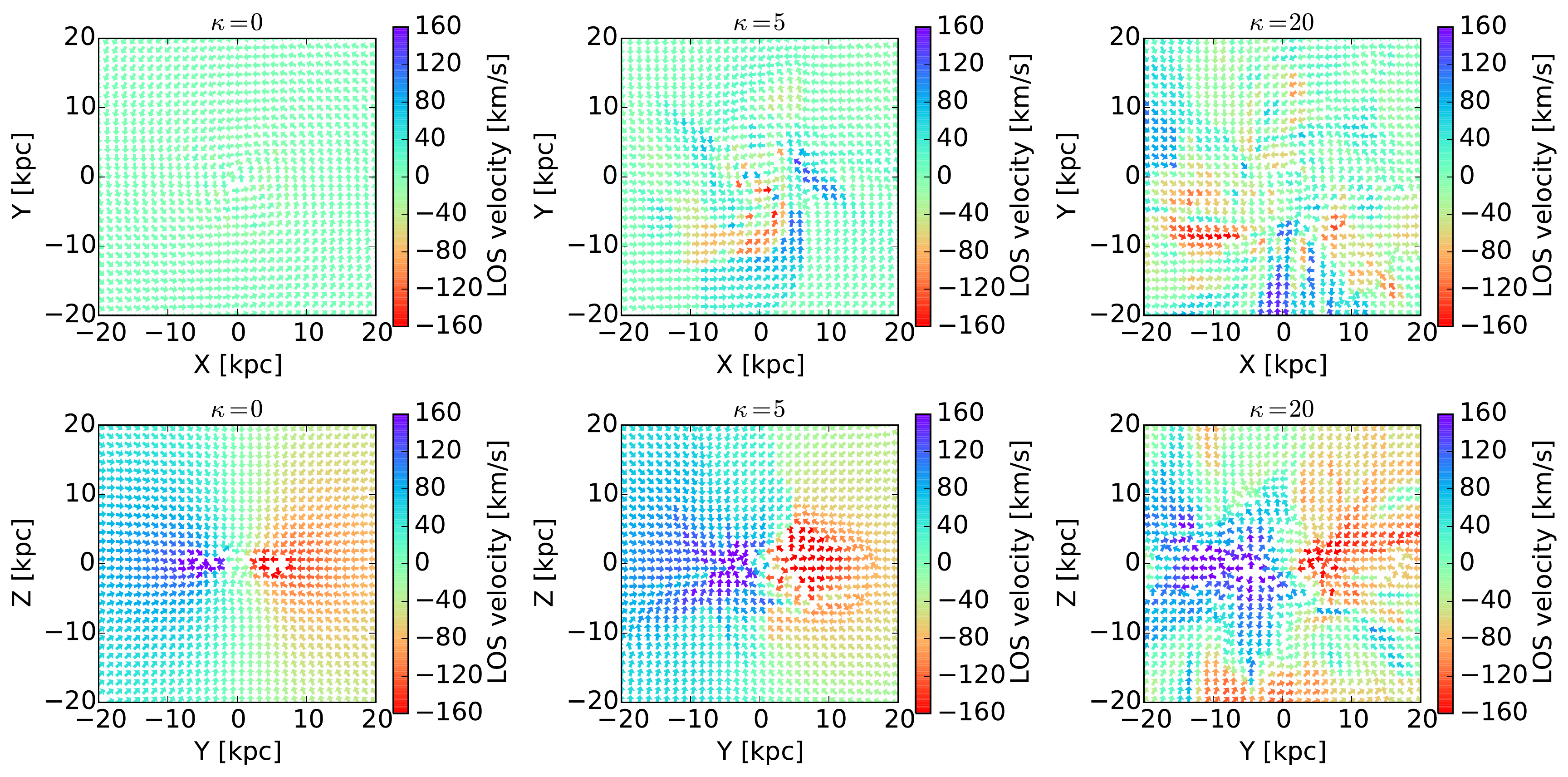}
\caption{Maps of the mass-weighted average velocity field corresponding to the density projections of \autoref{fig:mwmaps-rho}. Arrows show direction (but not strength) of the mean velocity field perpendicular to the line of sight. Colours indicate the line-of-sight component strength and direction where blue means approaching and red means receding from the observer. While the velocity is dominated by global rotation when feedback is weak (left, SN feedback only), the gas motion develops additional small-scale motion and flows vertical to the disk (middle) until it becomes almost isotropic turbulence (right).}
\label{fig:mwmaps-vel}
\end{figure*}

\begin{figure}
\centering
\includegraphics[width=\columnwidth]{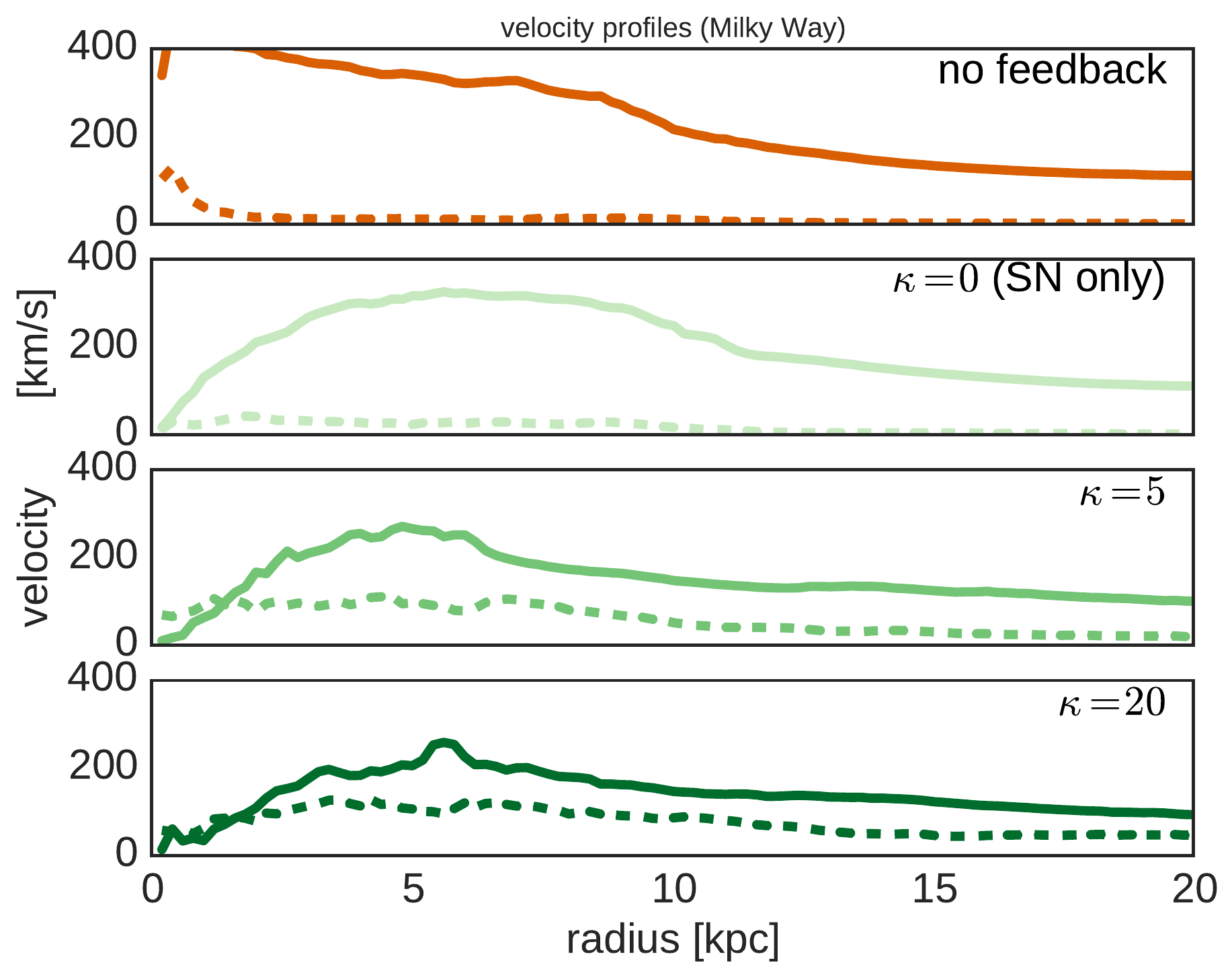}
\caption{Velocity profiles of  circular velocity and vertical velocity dispersion analogous to the one in \autoref{fig:velprofdw}, but for the Milky Way simulations with increasing values of the dust opacity $\kappa$ from top to bottom. While the rotational velocity $\overline{V_\theta}$ (solid lines) becomes smaller, the vertical velocity dispersion $\sigma_z$ (dashed lines) increases.}
\label{fig:velprofmw}
\end{figure}

Following the methodology explored for the first time in \cite{2014MNRAS.444.2837R}, we now consider a feedback model based on radiation from young stars efficiently absorbed by dust, and converted into kinetic energy through the infrared radiation force. Bear in mind that this model has been designed to be very optimistic, in order to maximise the effect of 
the radiation pressure on dust grains. Realistic modeling of the physical underlying processes is not our main objective here and we rather want to obtain an efficient feedback mechanism, launching a strong enough galactic wind, and analyse the possible effect of the resulting fountain flow on a magnetic dynamo. Images of gas density projections are shown in \autoref{fig:mwmaps-rho}. We have modelled our Milky-Way-like galaxy using two different values for the dust opacity: 
$\kappa=5~{\rm cm}^2/{\rm g}$ and $\kappa=20~{\rm cm}^2/{\rm g}$, which span a range of realistic dust temperatures. The latter model gives rise to the strongest galactic fountain, and resembles in many aspects to the dwarf galaxy case with supernovae feedback. The former, lower opacity case appears as less energetic, with a weaker wind and slightly smaller turbulence. These differences can be seen in the gas images in \autoref{fig:mwmaps-vel}, and are expressed quantitatively using the vertical velocity dispersion and the rotational velocity profiles (see \autoref{fig:velprofmw}), the high opacity, more extreme case giving rise to a quasi-spheroidal galaxy, with $V/\sigma \simeq 2$. 

\begin{figure}
\centering
\includegraphics[width=\columnwidth]{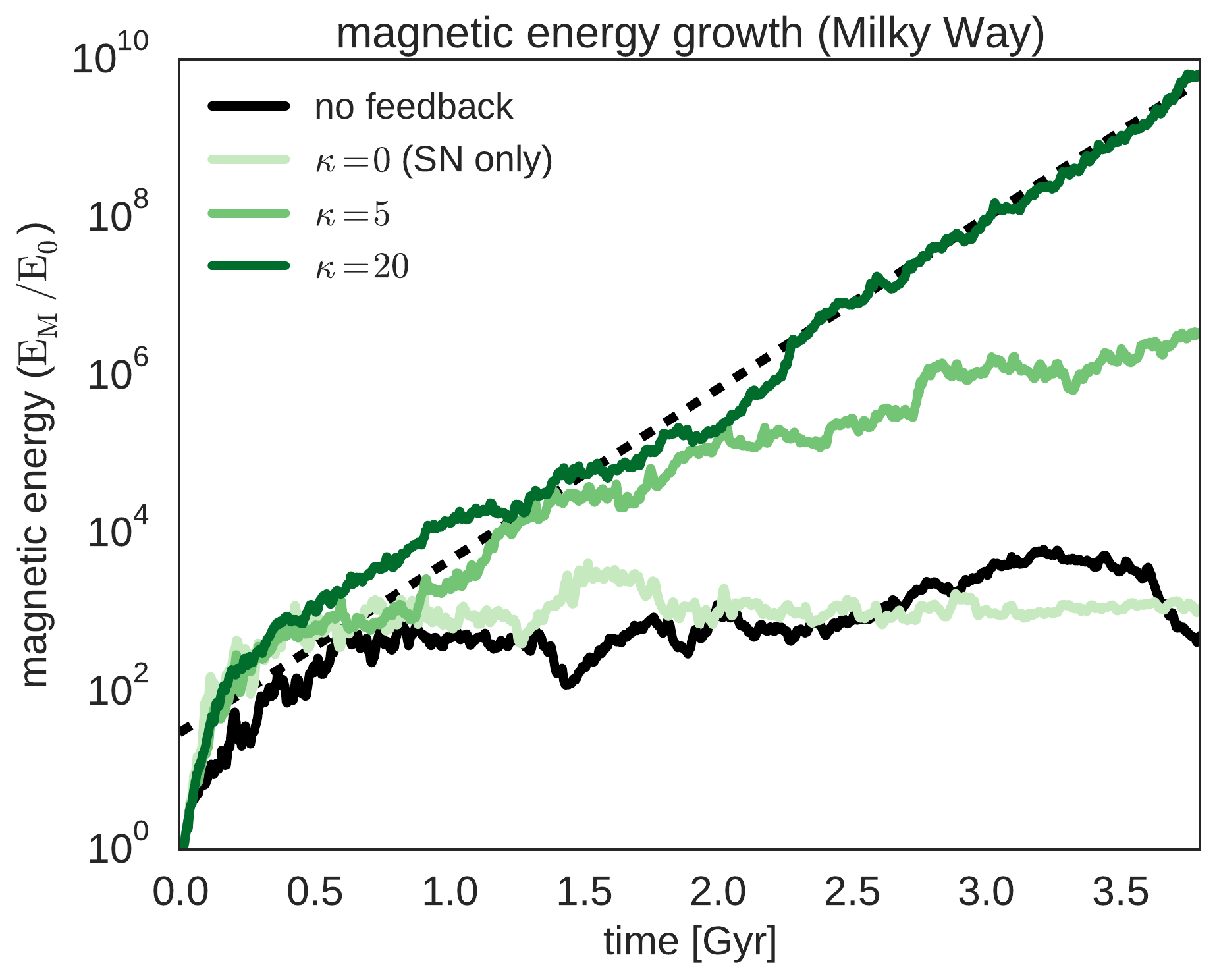}
\caption{Magnetic energy time evolution in the Milky Way simulations, normalised to their initial values. The dashed line marks an exponential growth $\exp{(\Gamma t)}$ at a rate of $\Gamma$ = 5 Gyr$^{-1}$ for comparison. Without feedback or only supernova feedback ($\kappa=0$), the energy does not grow after the initial collapse. Increasing the effective radiative feedback strength $\kappa$ boosts the growth rate $\Gamma$.}
\label{fig:mwenergy}
\end{figure}

When looking at the corresponding magnetic energy evolution on \autoref{fig:mwenergy}, it is interesting to notice that the higher velocity dispersion corresponds to the faster growth rate. Although both galaxies show an exponential growth of the magnetic energy, only the high opacity simulation reaches a growth rate as high as $\Gamma \simeq \Omega$. The low opacity case only reaches $\Gamma \simeq \Omega/2$. This is again in line with the theory of small-scale magnetic dynamos, for which the growth rate $\Gamma \simeq v(\ell)/\ell$, is proportional to the amplitude of the velocity fluctuations. Here again, our turbulent forcing scale is the size of the entire galaxy, and the dynamo growth rate is set by the the dissipation scale, which in our case corresponds to the numerical spatial resolution.

\begin{figure*}
\centering
\includegraphics[width=\linewidth]{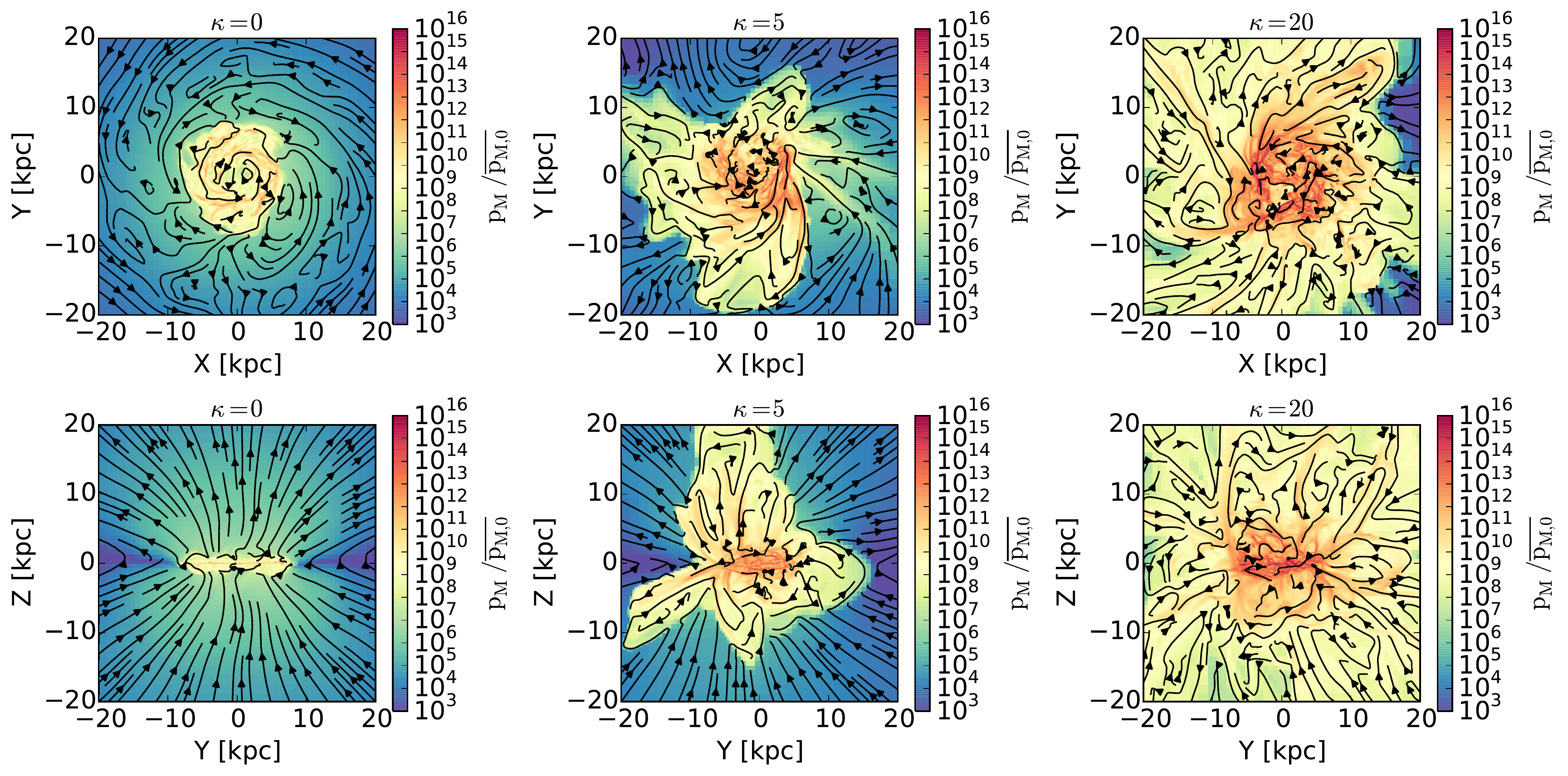}
\caption{Maps of the mass-weighted magnetic pressure corresponding to the density projections of \autoref{fig:mwmaps-rho} of Milky Way simulations with initial dipole field, normalised to the initial average magnetic pressure. Overlaid in black are streamlines of the mean field perpendicular to the line of sight. With weak feedback (left), the magnetic field retains the initial large-scale dipole structure outside the disk and a toroidal field inside the disk. Increasing feedback (mid and right) makes the field more small-scale.}
\label{fig:mwmaps-mag}
\end{figure*}

\section{Discussion}
\label{chap:discussion}

We have performed MHD simulations of feedback-dominated galaxies, both dwarf galaxies and  Milky-Way-sized galaxies. We have shown that, if feedback processes are strong enough, a small-scale dynamo sets in, with a typical kinetic energy injection scale $L$ corresponding to the size of the entire galaxy (which is in our case is close to the halo scale radius $r_s$), and a typical magnetic dissipation scale $\ell$ corresponding in our case to the adopted spatial resolution. 
We have observed an exponential increase of the magnetic energy, with growth rate $\Gamma \ge \Omega$, higher than the galaxy rotation rate, possibly much higher if one considers realistic microscopic diffusion processes instead of only numerical diffusion. 

Three important aspects are missing in order to apply our findings to the origin of galactic magnetic fields: 1-~We have considered rather idealised simulations of galaxies in isolation. 2-~We have only described the kinematic phase, deferring the discussion of the saturation to a companion paper (Paper~II). 3-~We have considered feedback-dominated galaxies, which are relevant for the high-redshift universe. What will happen after this feedback-dominated epoch, for quiescent, razor-thin galactic discs ? In this section, we speculate on possible cosmological consequences of our results on the nature of the magnetic field in high redshift galaxies, as well as the magnetic strength and topology in lower redshift galaxies.

\subsection{Cosmological implications for high redshift galaxies}

Although our numerical simulations were not performed in a realistic cosmological context, we can still draw conclusions for the cosmic evolution of magnetic fields, assuming that the universe is made of a collection of halos of various masses, and generalised our results using simple analytical estimates. For this purpose, we will assume that high-redshift galaxies are all dominated by efficient feedback processes, so that galactic winds can drive a powerful fountain, resulting in a gas rich, turbulence-dominated corona, with a size equal to $r_s$, the halo scale radius, that sets the turbulence injection scale. As shown in the previous sections, a very efficient small-scale dynamo is likely to develop, with a growth rate larger (possibly much larger) than the rotation rate of the galaxy. Although a detailed study of the saturation phase is required to study how fast the field will reach equipartition (and at what scales, see Paper~II), we postulate here that each halo reaches equipartition between magnetic and turbulent kinetic energy almost instantaneously, within a volume set by the halo scale radius $r_s$. This leads to the equipartition value for the field:
\begin{equation}
\frac{B_{\rm eq}^2}{4\pi} = \rho_{\rm gas} \sigma_{\rm turb}^2 
\end{equation}
For the gas density, we assume that its average value can be approximate by the baryon fraction $f_b = \Omega_b / \Omega_m$ of the total mass halo scale density
\begin{equation}
\rho_{\rm gas} \simeq f_b \rho_s
\end{equation}
and for the turbulence velocity dispersion, we adopt the value inferred by our simulations, namely the halo maximum circular velocity (at $r_s$)
\begin{equation}
\sigma_{\rm turb}^2 \simeq V_{\rm max}^2 \simeq  0.193 \times 4\pi G \rho_s r_s^2
\end{equation}
Using standard redshift-dependant functions for these two halo parameters $r_s$ and $\rho_s$, assuming an average halo concentration parameter $c=8$
and a baryon fraction $f_b \simeq 0.18$, we obtain a prediction for the saturated field
\begin{equation}
B_{\rm eq} \simeq 3\mu G \left( 1+z \right)^2 \left( \frac{M_{200}}{10^{10}M_{\odot}}\right)^{1/3}
\end{equation}
The resulting magnetic pressure scales as a function of mass and redshift exactly like a ``viral pressure'' in the halo.
Note that this equipartition field increases quite fast with increasing redshift, but quite slowly with increasing halo mass.
For a feedback-dominated, Milky-Way-sized galaxy, at redshift $z=2$, one still predicts a rather strong magnetic field, slightly above 100$\mu G$. 
Although very interesting in setting the foundations for a theory of small-scale dynamo in the cosmological context, our present discussion remains speculative, in the sense that we do not properly model cosmological infall and the associated hierarchical merging of smaller structures. This could lead to a dilution of the dynamo-amplified field and lower the growth rate. It is therefore of primary importance to simulate such feedback-dominated galaxy formation models with both MHD and a realistic cosmological environment.

\subsection{Transition to quiescent, low redshift galaxies}

\begin{figure*}
\centering
\includegraphics[width=\linewidth]{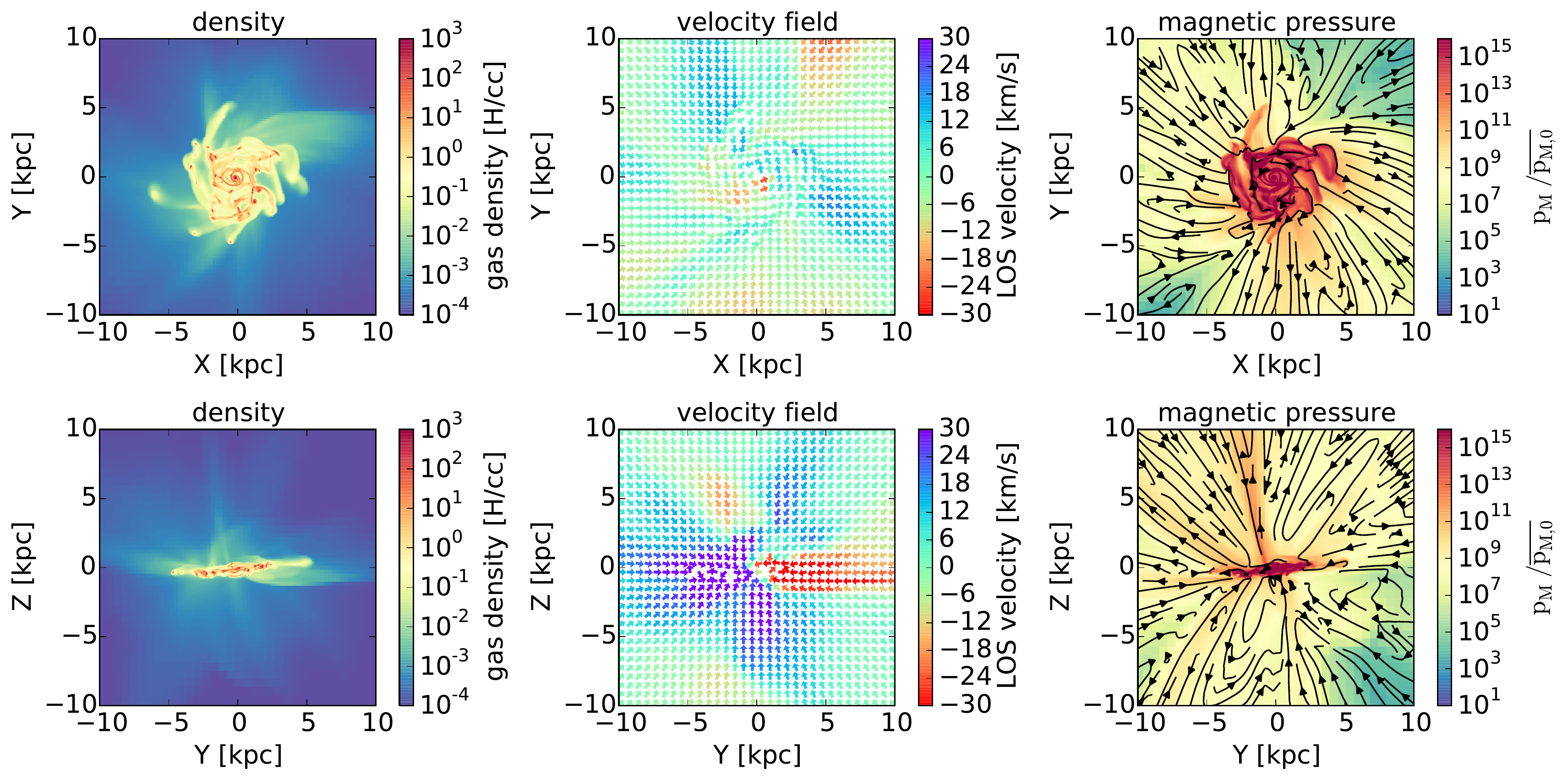}
\caption{Face-on (top row) and side-on (bottom row) views of mass-weighted projections of density (left), velocity fields (middle) and magnetic field (right) of suppressed feedback simulation at 3.8 Gyr. After feedback was switched off, the gas collapses into a thin clumpy disk. Turbulence becomes weaker, and the magnetic pressure is compressed in the disk with large-scale field lines.}
\label{fig:dwarf_sup_maps}
\end{figure*}

The cosmological applications we have derived from a simple analytical extension of our numerical results was within the context of feedback-dominated, high-redshift galaxies. Low redshift galaxies are however quite different. We see in our nearby universe many grand design, quiescent disk galaxies, with very thin, low velocity dispersion disks, and for larger halo masses, we even see red and dead elliptical galaxies. The present day universe is therefore very quiet, and strong feedback effects are absent, except may be in some very intense starbursts, usually triggered by (very rare) merger events. Our present methodology, based on isolated, gas rich, cooling halos, does not allow to explore the low redshift universe self-consistently, unless one artificially switches off feedback processes. This is the strategy we adopt in this section, in order to explore the consequence of evolving our simulated objects from a feedback-dominated state, to a more quiescent state, without strong galactic fountains, resulting in much thinner, rotationally supported disks. 

It is indeed very important to estimate how the magnetic field, amplified first through a turbulent-driven small-scale dynamo, could evolve into a large-scale field,
mostly driven by rotational shear. The obvious question one might ask is: Does the magnetic energy disappear, as the small-scale turbulent field reconnects on small scales? 
How intense will the surviving large-scale and mostly toroidal magnetic field be, after the galaxy develops into a thin, centrifugally supposed disk? For this reason, we decide to re-start our dwarf galaxy simulations after 3~Gyr of small-scale dynamo amplification, but without any stellar feedback. In the absence of galactic winds, the turbulent corona rapidly collapses back to a thin disk, and a mostly toroidal field appears after a few rotations. Images for gas density, velocity field and magnetic field of the galaxy after it has collapsed are shown in \autoref{fig:dwarf_sup_maps}.

\begin{figure}
\centering
\includegraphics[width=\columnwidth]{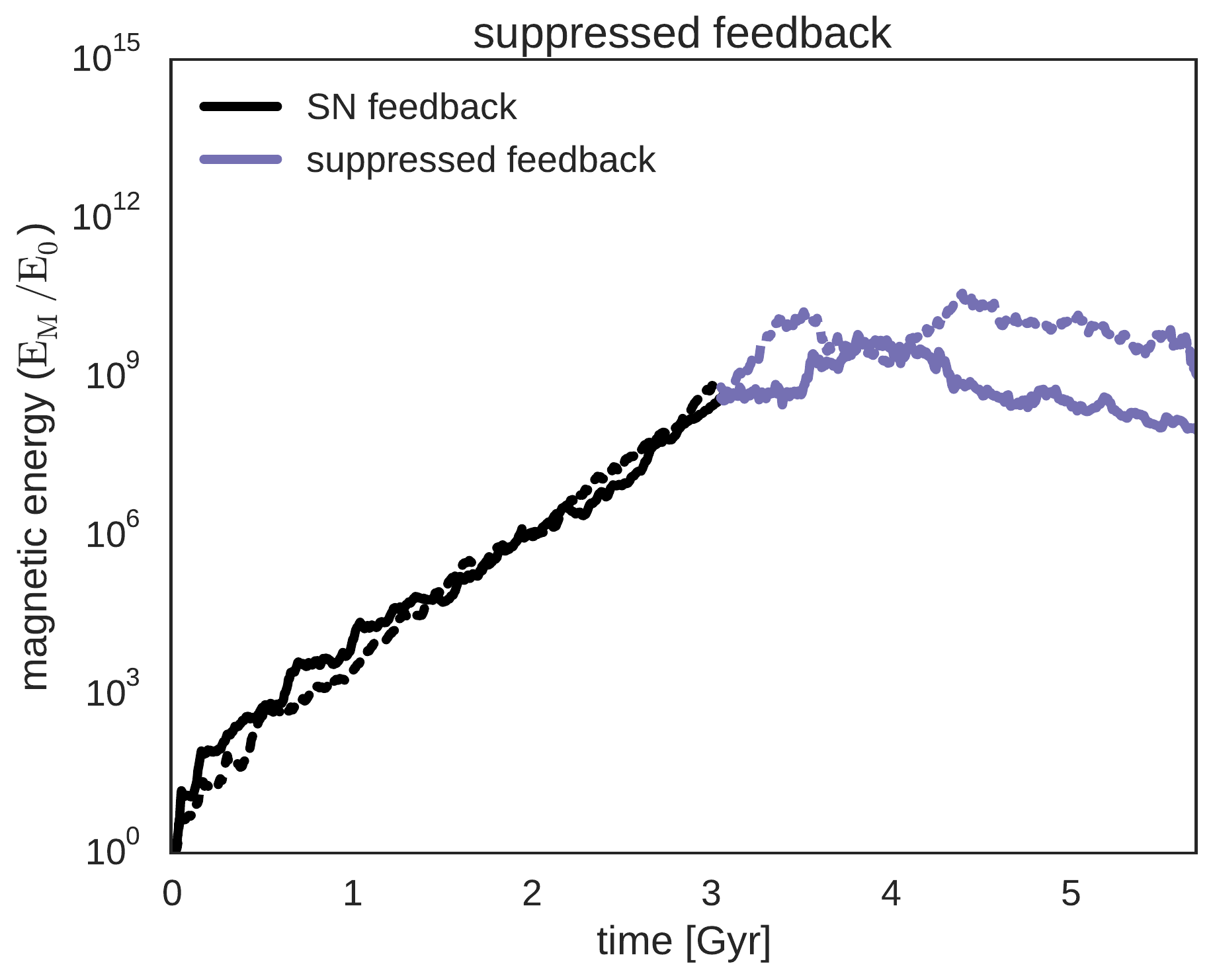}
\caption{Time evolution of the magnetic energy inside the simulation box of dwarf galaxies with feedback (black) and re-runs of the same simulations from 3 Gyr, but with feedback switched off (blue), with initial dipole (solid) and quadrupole (dashed) magnetic field configuration. When feedback becomes suppressed, the exponential growth is halted and the total magnetic energy remains roughly constant.}
\label{fig:energy_dwarf_sup}
\end{figure}

We see in \autoref{fig:energy_dwarf_sup} the evolution of the magnetic energy versus time for this ``suppressed feedback'' simulation. We see that overall, after the thin disk appears, the magnetic energy is for the most part conserved. Two competing effects are indeed at work here: gravitational collapse that amplifies the field from a corona-diluted state to a thin disk-concentrated state, on one hand, and magnetic losses due to reconnection of mostly small-scale field lines in the mid plane of the disk. What is interesting and highly non-trivial, is that these two effects basically cancel each other, and that the final magnetic energy in the thin disk is the same than the initial magnetic energy in the large, turbulence-supported corona. 

Large-scale, collapse-amplified magnetic fields have replaced small-scale, reconnection-suppressed fluctuations. Large-scale modes arise in the small-scale dynamo picture because the field is amplified on all scales at the same rate, up to the turbulence forcing scale, a well-known property of small-scale dynamos. Enough magnetic energy has been stored on large scales, so it can survive and compensate for reconnection and field cancellation effects during the collapse.

We would like to stress here again that this new type of simulations with suppressed feedback was performed in the weak field, pure kinematic regime as well. Since we believe that after the feedback-driven, small-scale dynamo phase, the field has probably quickly reached equipartition, we need to study this transition from thick-corona to thin-disk using a properly saturated field, which is the purpose of paper~II.

\begin{figure}
\centering
\includegraphics[width=\columnwidth]{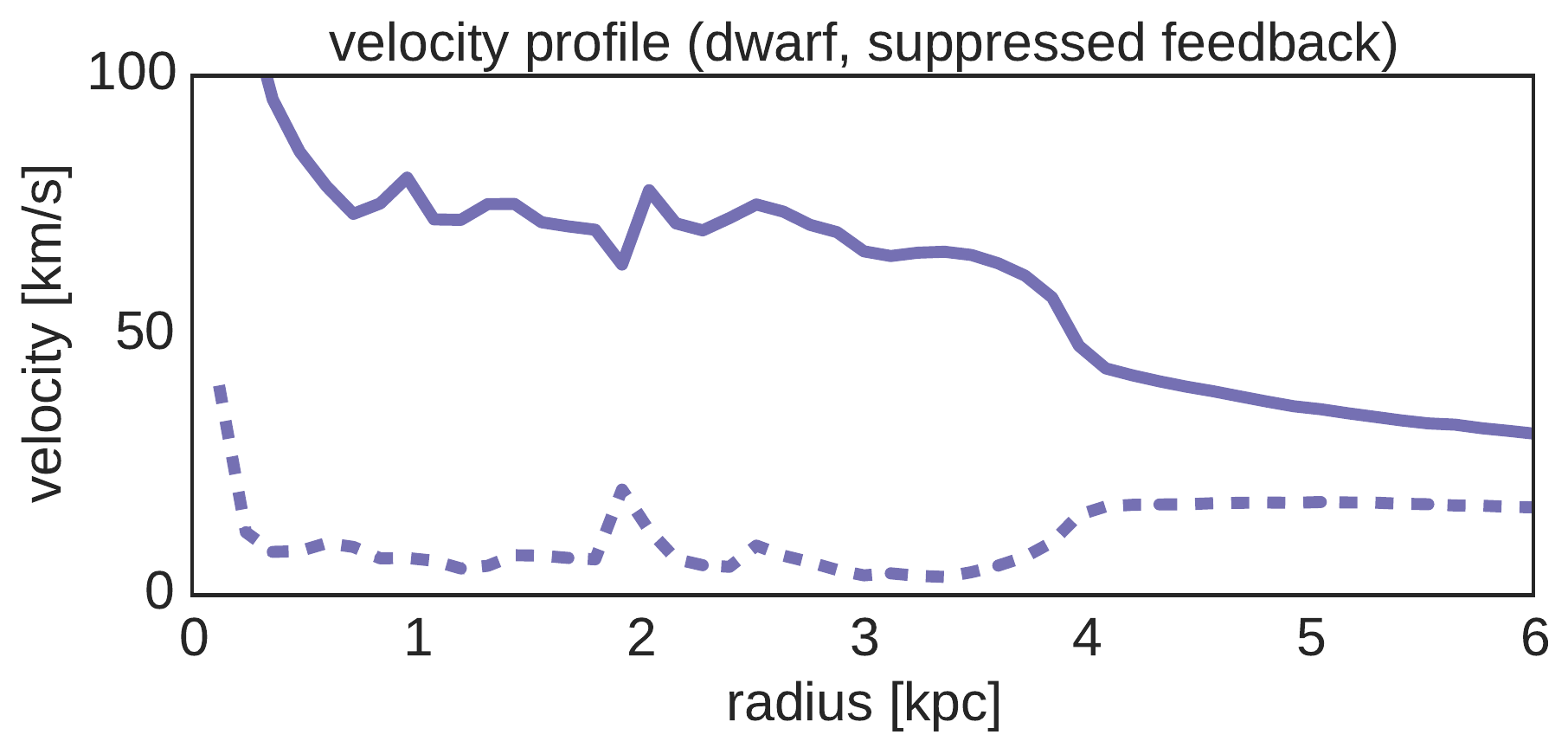}
\caption{Velocity profiles of circular velocity and vertical velocity dispersion in the suppressed feedback run, as in \autoref{fig:velprofdw}, 1~Gyr after the feedback has been switched off. Gas motions go back to being rotation-dominated, as in the no-feedback case.}
\label{fig:vel_prof_sup}
\end{figure}

\begin{figure}
\centering
\includegraphics[width=\columnwidth]{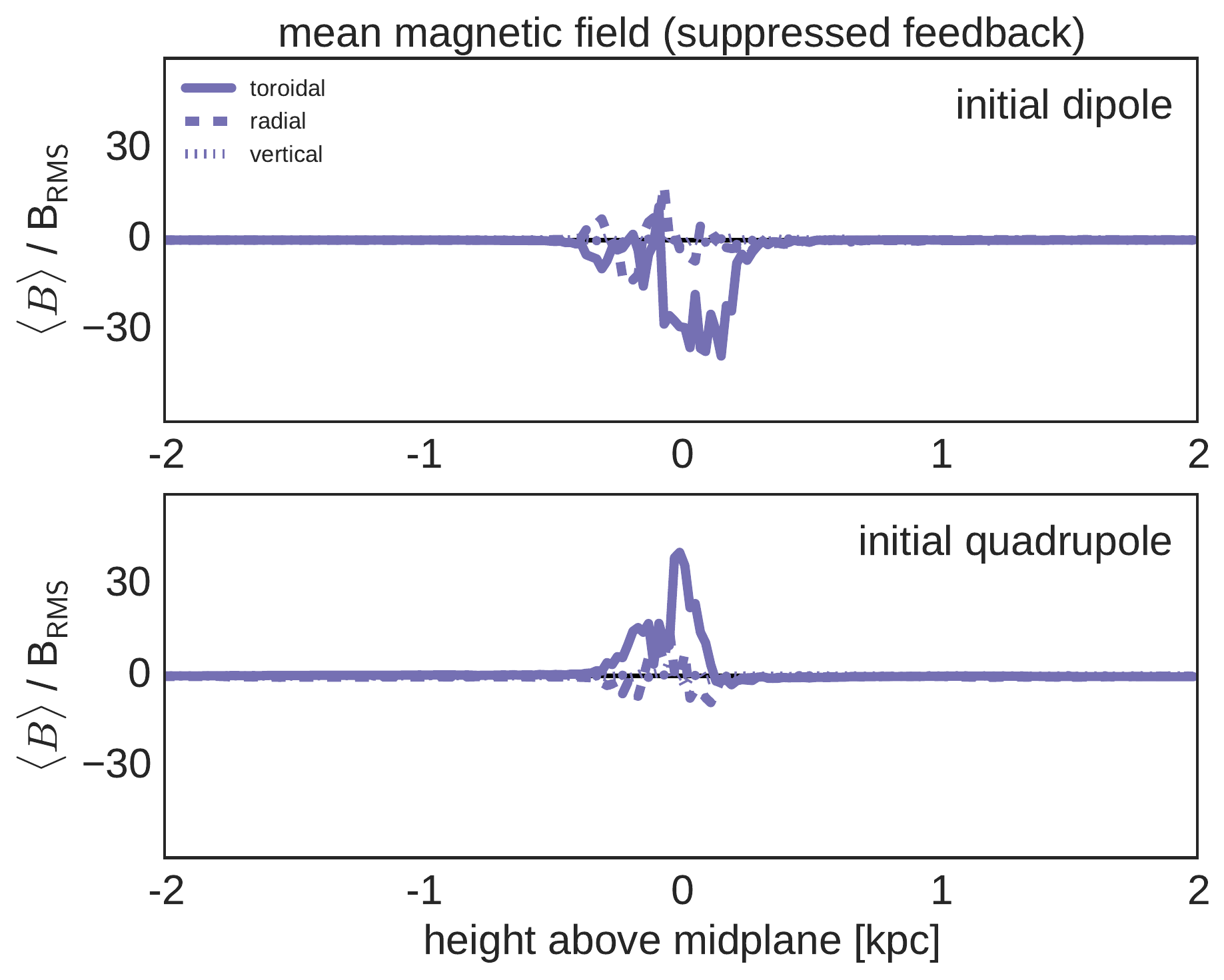}
\caption{Average magnetic field components after feedback has been switched off, versus the vertical height relative to the galactic mid plane, computed as in \autoref{fig:dwnofbk-midplane} and normalized to $\upright{B}\subscript{RMS}$. After collapse, the magnetic field is dominated by its toroidal component, but we also see a non-negligible radial component. The toroidal component has also developed an even symmetry across the mid plane, as in the simulations with initial quadrupole and without feedback, even if the initial condition was a dipole (top).}
\label{fig:parity_sup}
\end{figure}

Nevertheless, we can also study the topology of the field after the disk has settled in a thin, rotation-supported state. \autoref{fig:parity_sup} shows the toroidal, radial and vertical field in the disk 2~Gyr after the feedback has been suppressed. The field is mostly tangential, with however stronger radial and vertical components, compared to the no-feedback case. This is in agreement with observations which show pitch angles as high as $30^\circ$ \citep{patrikeev2006analysis}. The most interesting results is the topology of the field, which appears quadrupole-like, even if we start with only a dipole in the initial conditions. Dipole-like modes present in the large-scale magnetic field of the corona have odd parity. They will cancel in the mid plane after the gas has collapsed into a thin disk (like in the no-feedback case). Quadrupole-like modes, on the other hand, have even parity and they will be combined non-destructively in the mid plane after the collapse of the turbulent corona. One interesting prediction of the scenario which we consider here is therefore a systematic quadrupole field topology, independently of the initial topology of the primordial field. This result is in very good agreement with observational data of the Milky Way \citep{taylor2009rotation,oppermann2012improved}, as well as external galaxies \citep{braun2010westerbork,mao2012magnetic}.

\section{Conclusions}
\label{chap:conclusions}

We have performed MHD simulations of cooling halos, for both dwarf and Milky-Way-sized haloes, in the kinematic regime where magnetic field are weak enough so that the effect of the Lorentz force on the turbulent flow is insignificant. Using supernovae feedback for dwarf galaxies, as well as radiation feedback for large galaxies, we have shown that a small-scale dynamo quickly sets in, with the turbulent energy injection scale roughly equal to the halo scale radius $r_s$, and dissipation scale roughly equal to 4-5 cell sizes. In agreement with small-scale dynamo theory, we observe an exponential amplification of the magnetic energy on all scales, up to the injection scale, with a growth rate at least equal to twice the disk rotation 
rate. The growth rate in the kinematic phase is set by the adopted numerical resolution. In our simulated halo, we need to have a resolution of 20~pc or better (for $r_s \simeq 3.5$~kpc) to obtain a significant growth rate (as high as $\Gamma \simeq \Omega$). 

This corresponds roughly to 100 resolution elements per turbulent energy injection scale. We believe this resolution corresponds to the critical magnetic Reynolds number $R_{m} \simeq 35$, beyond which magnetic field amplification through small-scale dynamo is possible. Note that for weaker feedback scenarios, the kinetic energy injection scale is likely to be smaller, which translates into more stringent resolution requirements. This analysis should be kept in mind, when one wants to extend this to the cosmological context. One difficulty might arise when one considers infall of pristine, low magnetic field gas as a source of field dilution. This could require a larger small-scale dynamo growth rate for the field to be able to increase, and therefore results in even more demanding resolution requirements. 

We have shown that a small-scale dynamo, driven by strong feedback processes in high-redshift galaxies, could be the origin of galactic magnetic fields. This scenario is completely different than the more traditional large-scale dynamo approach, in the sense that the small-scale dynamo acts very quickly at amplifying the field up to the turbulent injection scale. The new ingredient is here the fact that feedback processes at high redshift are probably energetic enough, so that this injection scale is the size of the entire galaxy (more precisely the halo scale radius $r_s$). We therefore have a small-scale dynamo, together with a large-scale forcing, hence enabling the fast amplification of the field all the way up to the scale of the entire galaxy. 

We have also shown, using simple numerical experiments with suppressed feedback, that the field can evolve into a large-scale toroidal, quadrupole field in a low-redshift quiescent state, although it has been amplified by a small-scale dynamo. Compared to the large-scale dynamo picture, for which the field is amplified in a thin disk over several Gyr, the small-scale dynamo picture completely reverses the point of view, with a very strong field already in place at high redshift (after a feedback-dominated epoch) that slowly evolves until the present epoch. This large-scale field could even slowly decay, on resistive time scales, and still be large enough to account for the observed field strength in nearby galaxies. In this new picture, large-scale dynamos are not required anymore to amplify the field from its primordial value around $10^{-20} G$ to $\mu G$ levels. Instead, they are needed to maintain the field at a roughly constant level by compensating dissipative losses.

\section*{Acknowledgements}

We thank the anonymous referee for their thoughtful comments and helpful suggestions. This work was funded by the Swiss National Science Foundation SNF. All Computations were performed on the local zBox super computer and on the Piz Dora cluster at the Swiss National Supercomputing Center CSCS.

\bibliography{romain,michael}

\end{document}